\documentclass[pra,aps,reprint,a4paper,superscriptaddress,floatfix]{revtex4-2}
\usepackage{times}
\usepackage{graphicx}
\usepackage{epsfig}
\usepackage{amsfonts}
\usepackage{amsmath}
\usepackage{amssymb}
\usepackage{dsfont}
\usepackage{amsthm}
\usepackage{color,xcolor,ulem}
\usepackage[colorlinks=true,linkcolor=blue,citecolor=blue,urlcolor=blue,breaklinks]{hyperref}
\usepackage{comment}
\usepackage{braket}
\usepackage{physics}
\usepackage[breaklinks]{hyperref}
\usepackage{bbm}

\usepackage{float}

\usepackage{braket}

\begin{document}
	
	
\title{Experimental demonstration of the equivalence of entropic uncertainty with wave-particle duality}


\author{Daniel Spegel-Lexne}
\email{These authors have contributed equally to this work.}
\affiliation{Institutionen f\"{o}r Systemteknik, Link\"opings Universitet, 581 83 Link\"oping, Sweden}

\author{Santiago G\'omez}
\email{These authors have contributed equally to this work.}
\affiliation{Departamento de Física, Universidad de Concepción, Casilla 160‑C, Concepción, Chile.}
\affiliation{Millennium Institute for Research in Optics, Universidad de Concepción, Casilla 160‑C, Concepción, Chile.}

\author{Joakim Argillander}
\affiliation{Institutionen f\"{o}r Systemteknik, Link\"opings Universitet, 581 83 Link\"oping, Sweden}

\author{Marcin Pawłowski}
\affiliation{International Centre for Theory of Quantum Technologies,
University of Gdańsk, Jana Bazynskiego 8, 80-309 Gdańsk, Poland}

\author{Pedro R. Dieguez}
\affiliation{International Centre for Theory of Quantum Technologies,
University of Gdańsk, Jana Bazynskiego 8, 80-309 Gdańsk, Poland}

\author{Alvaro Alarc\'on}
\affiliation{Institutionen f\"{o}r Systemteknik, Link\"opings Universitet, 581 83 Link\"oping, Sweden}

\author{Guilherme B. Xavier}
\email{guilherme.b.xavier@liu.se}
	\affiliation{Institutionen f\"{o}r Systemteknik, Link\"opings Universitet, 581 83 Link\"oping, Sweden}

\begin{abstract}
Wave-particle duality is one of the most striking and counter-intuitive features of quantum mechanics, illustrating that two incompatible observables cannot be measured simultaneously with arbitrary precision. In this work, we experimentally demonstrate the equivalence of wave-particle duality and entropic uncertainty relations using orbital angular momentum (OAM) states of light. Our experiment utilizes an innovative and reconfigurable platform composed of few-mode optical fibers and photonic lanterns, showcasing the versatility of this technology for quantum information processing. Our results provide fundamental insights into the complementarity principle from an informational perspective, with implications for the broader field of quantum technologies.
\end{abstract}
\maketitle

\section{Introduction}

\begin{figure*}[!ht]
\centering
\includegraphics[width=0.8\textwidth]{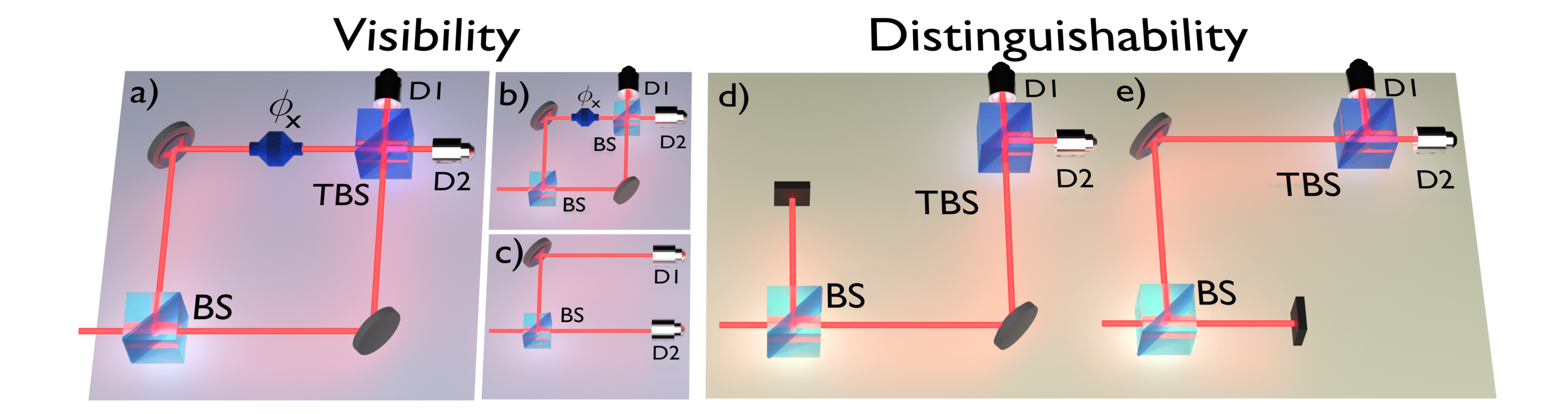}
\caption{Different experimental configurations for complementarity measurements. a) Mach-Zehnder (MZ) interferometer with a tunable beam splitter (TBS) to recombine the paths and a modulator applying a relative phase shift $\phi_x$ between the two arms. b) TBS adjusted to equal transmission and reflection coefficients, yielding full interferometric visibility. c) TBS adjusted to complete transmission or reflection, equivalent to it being removed. In this case, full path information is available; thus, no interference can be observed. d and e) For the distinguishability measurements, each path is individually blocked and the detection events are recorded for any setting of the TBS. In both cases, no interference pattern is possible.}
\label{Fig1}
\end{figure*}

The uncertainty principle states that some pairs of observables cannot be known simultaneously with high precision. On the other hand, 
Bohr's complementarity principle manifested its key role in modern physics by submitting matter and radiation to a unifying framework, where it is expected to exclusively behave either in a {\it wave-} or in a {\it particle-like} manner, depending on the peculiarities of an interferometric setup~\cite{dieguez2022experimental}. Interestingly, by examining the operational meaning of min- and max -entropies~\cite{konig2009operational,coles2017entropic}, the uncertainty and complementarity principles were unified for two-path interferometers~\cite{coles2014equivalence}. This unification was extended to multipath interferometers in Ref.~\cite{coles2016entropic} by considering wave and particle duality through two complementarity guessing games. This framework introduced a quantifier for fringe visibility applicable to multipath interferometers, revealing a proper trade-off with a generalized version of path-distinguishability.

 The entropic uncertainty relations (EURs) can be applied to interferometers in two different manners. One concerns preparation uncertainty, where a quantum state cannot be prepared with certainty for two complementary observables, while the other involves measurement uncertainty, indicating that two complementary observables cannot be jointly measured~\cite{coles2014equivalence}. Complementarity and uncertainty then pose a challenge to classical assumptions, such as physical realism~\cite{dieguez2018information,dieguez2022experimental}, by proposing that quantum systems may possess properties that only emerge once the entire physical context, including the system and the classical measuring apparatus, is finally settled~\cite{bohr1928quantum, bohr1935can,dieguez2022experimental}. Wave-particle duality relations (WPDRs) aim to provide operational meaning to the complementarity principle, as it can be employed to serve as the fundamental mechanism driving advantages over classical tasks~\cite{coles2017entropic}. The equivalence between EURs and WPDRs may find applications across various quantum technologies, including quantum communication~\cite{hameedi2017complementarity}, metrology~\cite{yadin2021metrological,len2022quantum}, cryptography~\cite{scarani2009security,koashi2009simple,berta2010uncertainty,mizutani2017information,zhang2023quantum}, and thermodynamics~\cite{pal2020experimental,koyuk2020thermodynamic,horowitz2020thermodynamic,falasco2020unifying,vieira2023exploring}.

Several experiments have been conducted to test the robustness of the complementarity principle, particularly through delayed-choice experiments~\cite{Wheeler,jackes2007experimental} and its generalizations, such as the entanglement-separability duality in bipartite systems~\cite{Peres00,Jennewein05}. This duality is examined in delayed-choice entanglement swapping experiments~\cite{Ma12} and quantum delayed-choice experiments~\cite{Terno,auccaise2012experimental,dieguez2022experimental}, which utilize quantum-controlled gates to create an effective superposition of a present and absent beam splitter within the experimental setup. Despite these efforts, debates persist~\cite{jackes2007experimental, dieguez2022experimental, vedovato2020extending, catani2021interference, catani2022aspects} regarding whether interference phenomena truly defy classical explanations or whether Bohr's complementarity principle needs updating~\cite{chrysosthemos2023updating}.

In this work, we present the first experimental demonstration of the equivalence between wave-particle duality and entropic uncertainty relations \cite{coles2014equivalence}, for which we rely on states encoded in the orbital angular momentum (OAM) degree of freedom of light \cite{Allen:92, Shen:19} to assess the entropic uncertainty both from the definition as well as using interferometric visibility and input distinguishability~\cite{coles2014equivalence}. The OAM of light is widely used in photonic quantum information due to its support for high-dimensional Hilbert spaces \cite{Erhard:18}. Traditionally, OAM-based experiments relied on bulk optics and free-space channels \cite{Leach:02, Karimi:09, Wang:12, Vallone2014, Krenn2014, Mirhosseini:15, Sit2017, Liu2020} due to the challenge of manipulating and transmitting these states over optical fibers, crucial for fiber-optic telecommunication compatibility.
Multi-mode fibers, although capable of supporting OAM modes, require complex compensation systems for stable transmission due to numerous transverse spatial modes \cite{Choi2012, Caramazza2019}. Recently, the telecom industry developed spatial division multiplexing (SDM) fibers, which support fewer transverse spatial modes to enhance transmission capacity \cite{Richardson2013}, and have become vital for processing and transmitting photonic transverse spatial quantum states \cite{Xavier:20}. Few-mode fibers, a type of SDM fiber, have successfully transmitted photonic OAM quantum states over long distances \cite{Cao2020, Alarcon2021} by linearly decomposing OAM modes into the linearly polarized (LP) modes supported by the fibers.
Furthermore, using photonic lanterns \cite{Birks:15} in an interferometric setup, we can dynamically excite different OAM modes in a few-mode fiber with ultra-fast response times \cite{Alarcon2023_2}. This method also enables ultra-fast reconfigurable projective measurements on OAM states \cite{Alarcon2023}. Employing these techniques we perform here entropic uncertainty measurements on the wave and particle aspects of OAM states using a few-mode fiber interferometric setup combined with a fiber-optical tunable beam splitter based on a Sagnac interferometer. Our setup can dynamically change the measurement operator such that the two complementarity extremes can be measured, as well as any value in between. We are able to experimentally verify the entropic uncertainty relations revealing the operational meaning of the complementarity principle, as well as opening new applications for quantum information processing.

\section{Preliminaries: EURs and WPDRs}

Wave and particle behaviors are quantified by the knowledge related to mutually unbiased observables \cite{coles2014equivalence,coles2016entropic,coles2017entropic} within the entropic uncertainty framework~\cite{coles2014equivalence}. The related WPDR  relies on the operational connection between the min- and max-entropies, with guessing probabilities. The wave-particle duality emerges in this context as a fundamental bound for two complementarity guessing games~\cite{coles2017entropic}. In the following, we briefly introduce the framework developed in Refs.~\cite{coles2014equivalence,coles2016entropic}, to present our experimental confirmation of the WPDR and EUR equivalence. For instance, for a classical-quantum state in the form of $\rho_{\mathcal{AB}}=\sum_{a}p_a\ket{a}\bra{a}\otimes \rho_{\mathcal{B}}^a$, and setting $Z$ as the which-path random variable associated with the basis $\mathbbm{Z}=\{\ket{z}\bra{z}\}$ of an $n$-dimensional Hilbert space $\mathcal{H_A}$, and $W$ as the random variable associated with a mutually unbiased basis $\mathbbm{W}$ in respect to $\mathbbm{Z}$, the following optimized wave and particle entropic uncertainty relation
\begin{equation}
\begin{aligned}
\label{eq: EUR}
& H_{\text{min}}(Z|B)+\min_{\mathbbm{W}} H_{\text{max}}(W)\geq \log_2{n},
\end{aligned}
\end{equation}
states that for a $n$-path interferometer, the sum of the ignorance about wave and particle behaviors is at least $\log_2{n}$ bits of information~\cite{coles2017entropic}. In the Appendix~\ref{Methods a} section, we detail the framework developed in Refs.~\cite{coles2014equivalence,coles2016entropic} that shows the equivalence between the EUR described in Eq.~\ref{eq: EUR} with a generalized path-distinguishability and interferometric visibility such that $\mathcal{D}^2 + \mathcal{V}^2 \leq 1$ holds~\cite{coles2016entropic}. 

\begin{figure*}[!t]
\centering
\includegraphics[width=18cm]{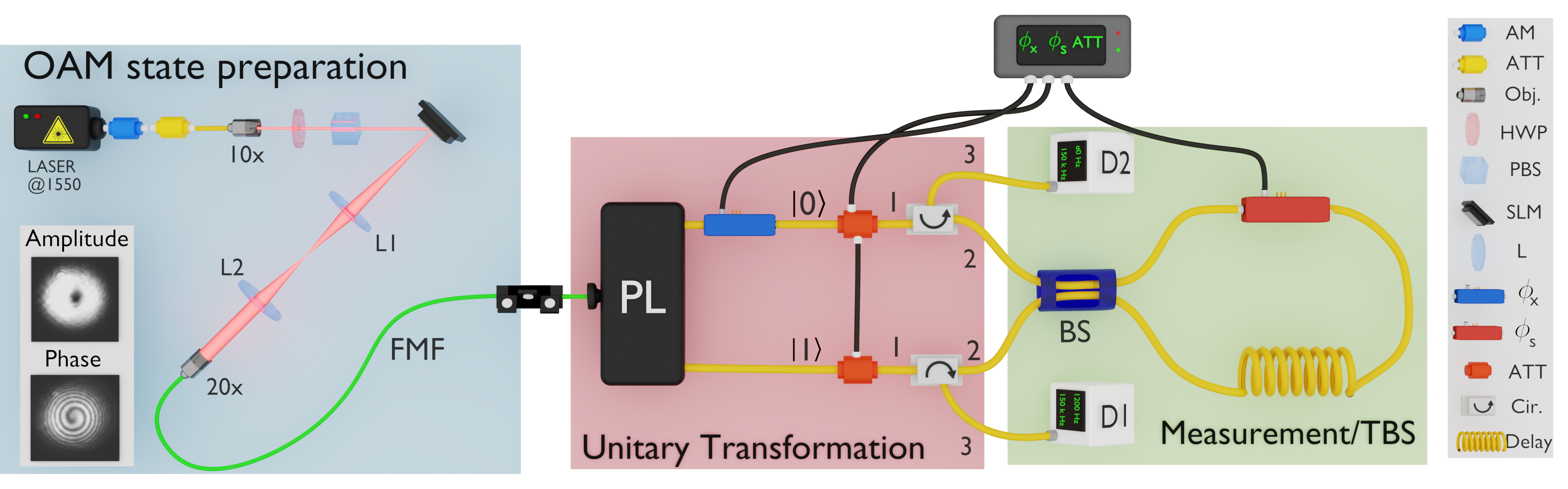}
\caption{The setup for probing the EUR over a $|\mathrm{OAM}_{+1}\rangle$ qubit is shown above.  The experiment consists of three main parts: the $|\mathrm{OAM}_{+1}\rangle$ mode source, the unitary transformation, and the measurement stage (or TBS). An amplitude modulator (AM) and an attenuator (ATT) adjust the average number of photons per pulse to $\mu = 0.2$. Additionally, the $|\mathrm{OAM}_{+1}\rangle$ state can be encoded by applying the appropriate forked diffraction grating to the spatial light modulator (SLM) and also using a 4f system (see Appendix \ref{Methods b} section). The amplitude and phase profiles of the OAM$_{+1}$ state are shown, following propagation through the few-mode fiber (FMF) with an InGaAs infrared camera with the laser source unattenuated. Then the unitary transformation stage performs a mapping from spatial to path information, yielding $\ket{\psi} = \frac{1}{\sqrt{2}}(\ket{0} + e^{i\phi_{x}}\ket{1})$, where a rotation to the state can be applied through the phase modulator $\phi_{x}$. The TBS is described by a Sagnac interferometer containing an optical delay and a phase modulator $\phi_{s}$, which controls the transmissivity and reflectivity of the BS, which determines the maximum obtained visibility.  On the other hand, to measure the distinguishabilities $\mathcal{D}_1$ and $\mathcal{D}_2$, it is necessary to block the paths before the Sagnac interferometer with electro-optical attenuators (ATT).}
\label{Fig2}
\end{figure*}
For binary interferometers, particle behavior is quantified by the knowledge of the observable $Z$ which indicates the path the system has taken inside the interferometer. In contrast, wave properties are often modeled as having a well-defined phase and being spatially delocalized. Hence, for a $2$-dimensional Hilbert space $\mathcal{H_A}$, wave properties are associated with the eigenstates of observables living in the $\text{X-Y}$ plane of the Bloch sphere, describing a set of observables that are mutually unbiased with $Z$.  Here, we employ an interferometric setting for which we will first refer to a standard Mach-Zehnder interferometer where the final beam splitter (BS), responsible for the recombination of the wave packets, is replaced with a tunable beam splitter (TBS) (Fig. \ref{Fig1}),
to test the equivalence of wave-particle duality with the following EUR
\begin{equation}
\begin{aligned}
   & H_{\text{min}}(Z)+\min_{\mathbbm{W}} H_{\text{max}}(W)\geq 1.
    \end{aligned}
\label{EUR1}
\end{equation}
The equivalence with a WPDR in our interferometric setting depicted in Fig.~\ref{Fig1}, which employs a 50:50 BS at the input, is obtained by demonstrating that min- and max- entropies are respectively related to particle and wave information of a quantum state  as \cite{coles2014equivalence}
\begin{equation}
\label{Eq6}
    H_{\text{min}}(Z)=-\log_2{\frac{1+\mathcal{D}}{2}},
    \end{equation}
where  the input distinguishability $\mathcal{D}$ can be experimentally obtained as the average
\begin{equation}
     \mathcal{D}:=\frac{1}{2}(\mathcal{D}_1+\mathcal{D}_2),
    \label{Eq2}
\end{equation} 
such that 
\begin{equation}
    \mathcal{D}_{1(2)}:=\left[\frac{\abs{p_1-p_2}}{p_1+p_2}\right]_{\text{path 1(2) blocked}},
    \end{equation}
and
\begin{equation}
    \min_{\mathbbm{W}} H_{\text{max}}(W)=\log_2{(1+\sqrt{1-\mathcal{V}^2})},
    \label{Eq7}
\end{equation}
with the usual interferometric visibility for a $2$-path interferometer defined as
\begin{equation}
    \mathcal{V}:=\frac{p_j^{\text{max}}-p_j^{\text{min}}}{p_j^{\text{max}}+p_j^{\text{min}}},
    \label{Eq1}
\end{equation}
where the maximization and minimization over the detection probability $p_j$ with $j=1,2$ are performed under the controllable phase of the interferometer $\phi_x$ (as depicted in Fig.~\ref{Fig2}). In this interferometric arrangement, a single photon interacts with a BS generating a path superposition state across the two paths, with a relative phase applied to one of the paths, creating the state $\ket{\psi}=\frac{1}{\sqrt{2}}(\ket{0}+ie^{i \phi_x}\ket{1})$, where $\ket{0}$ and $\ket{1}$ correspond to a path at the output of the 50:50 BS respectively, and the paths are recombined at a TBS (Fig. \ref{Fig1}a). The relative phase $\phi_x$ is added by a phase modulator, with $i$ being the imaginary component.

 A TBS has the capability of dynamically changing its transmission $(t)$ and reflection coefficients $(r)$, obeying the condition that $|r|^2 + |t|^2 = 1$. The interferometer has two orthogonal outputs where single-photon detectors D1 and D2 are placed. If the two paths are indistinguishable, then the probability amplitudes interfere at the second beamsplitter, with full interferometric visibility achieved if the TBS is set such that $r = t$ equivalent to a standard 50:50 BS (Fig. \ref{Fig1}b).  On the other hand, if the BS is completely removed ($t$ or $r = 1$), we have the situation in Fig. \ref{Fig1}c, in which the paths are completely distinguishable, and no interference behavior can be observed. It is also possible to obtain partial wave and particle behavior, effectively going through a continuum between the two extrema cases by adjusting the TBS.

 The visibility can be measured directly for different settings of the TBS from the detection probabilities at the outputs of the interferometer. Finally, in order to quantify the distinguishability, each path needs to be individually blocked (Figs. \ref{Fig1}d and e), and thus the probabilities can be measured for any state of the TBS. 

\section{Results}


Based on an all-in-fiber interferometer, we have implemented an experimental setup to test the EUR on an orbital angular momentum (OAM) quantum state. Instead of using a standard MZ interferometer, our setup is innovatively designed to observe the wave or particle behavior of an OAM state using SDM devices and a Sagnac interferometer as a TBS. The experimental setup, shown in Fig. \ref{Fig2}, can be divided into three main stages: state preparation, unitary transformation, and measurement (or tunable beam splitter). Let us present these three stages in sequence.
\par
A weak coherent state (WCS) is prepared using a continuous-wave telecom laser operating at $1550$ nm connected to a fiber-pigtailed amplitude modulator (AM) generating a pulse width of $40$ ns at a repetition rate of $150$ kHz, in series with an optical attenuator (ATT) to bring down the optical power to single-photon level. Next, a spatial light modulator (SLM) and a 4f system are used to prepare the $\ket{\mathrm{OAM}_{+1}}$ state (see Appendix \ref{Methods b} section). The light is coupled into a few-mode fiber (FMF) capable of supporting three spatial modes, the fundamental linearly polarized mode LP$_{01}$ and the two degenerate higher-order modes LP$_{11a}$ and LP$_{11b}$. Within the FMF, the OAM state is decomposed into LP components as follows $\ket{\mathrm{OAM}_{+1}} = \frac{1}{\sqrt{2}}(\ket{\mathrm{LP}_{11a}} + i\ket{\mathrm{LP}_{11b}}$ \cite{Alarcon2021}, where $i$ is the relative phase between the LP modes. We image the OAM$_{+1}$ mode after coupling to the FMF, by imaging its facet onto an InGaAS CCD infrared camera. The amplitude profile is imaged directly, while for the phase profile we interfere the OAM state with a Gaussian beam generated by the same laser and split before the SLM (not shown for simplicity) \cite{Alarcon2023_2}. This measurement is done with the laser without attenuation and in continuous wave mode and is shown in the inset of the OAM state preparation stage in Fig. \ref{Fig2}.
\par
When the OAM state reaches the unitary transformation stage, a 3-mode photonic lantern (PL) (Phoenix Photonics) is used as a spatial demultiplexer to perform a mapping from modal to path information as follows: $|$LP$_{11a}\rangle$ $\rightarrow |0\rangle$  and $|$LP$_{11b}\rangle$ $\rightarrow |1\rangle$, where $|0\rangle$ and $|1\rangle$ are the upper and lower arm of the interferometer respectively. The third port of the lantern, corresponding to the fundamental LP$_{01}$ mode, is not used. The two paths following the lantern mapping operation are single-mode optical fibers. A lithium niobate (LiNbO$_3$) fiber-pigtailed telecom phase modulator ($\phi_x$) is placed in the upper path, allowing unitary transformations on the superposition state $\frac{1}{\sqrt{2}}(\ket{0} + e^{i\phi_{x}}\ket{1})$. Manual polarization controllers (not shown for the sake of simplicity) are placed in each arm to optimize interference in the $50:50$ fiber beamsplitter (BS), as well as two variable electro-optical attenuators (Thorlabs V1550A) to control the transmissivity of each path.
\par
In transitioning from the unitary transformation stage to the measurement stage, each of the two paths passes through two optical circulators, which transmit each path component forward ($1 \rightarrow 2$, see Fig. \ref{Fig2}) to a $50:50$ fiber beamsplitter (BS), forming a fiber-optical Sagnac interferometer (SI). The SI operates as a tunable beam splitter controlled by a LiNbO$_3$ phase modulator ($\phi_s$). Following the BS, two wave packets are generated within the interferometer, traveling clockwise and counterclockwise. A $300$ m fiber optic delay line is used to ensure sufficient time separation for the phase modulation signal ($\phi_s$) to act on the wave packet propagating in only one of the internal directions, thus creating the change of relative phase necessary to generate tunability on the SI outputs \cite{Argillander_2022}. Within the SI there are two manual polarization controllers (also not shown for simplicity) to ensure that both wave packets recombine with the appropriate polarization in the BS, as well as aligning the polarization state at the phase modulator. After recombination, the wave packets are sent again to the circulators (now taking the direction $2 \rightarrow 3$) and are detected at D1 and D2. We employ one single-photon detector with a time multiplexing scheme to be able to measure both outputs simultaneously (Appendix \ref{Methods c} section). The detector is an InGaAs-based single-photon counting module (IdQuantique id210) running in gated-mode with $10\%$ overall detection efficiency, $3$ ns wide gate windows. Both the $\phi_s$ modulator and the detector are synchronized with the repetition rate of the source. The total loss of the measurement system is approximately $12$ dB, including the optical circulators. We adjust the input optical attenuator at the beginning of the setup to obtain $0.2$ photons on average per detection gate of $3$ ns just after the state is prepared by the unitary transformation operation. As such, a multi-photon probability of less than $2\%$ is achieved. 
\par
When $\phi_{s} = 0$, the SI behaves like a mirror, essentially equivalent to having no beam splitter installed at the measurement stage, as both inputs to the BS are reflected to the corresponding circulators. Despite continuously modulating $\phi_{x}$, no interference pattern (particle behavior) occurs, as deduced from  Eq. \ref{Eq9} and Eq. \ref{Eq10} in the Appendix~\ref{Methods a} section. On the other hand, when $\phi_{s} = \pi/2$, the SI operates as a beam splitter with equal transmission and reflection coefficients. Consequently, the measurement will exhibit interference fringes with maximum contrast (wave behavior). Any intermediate $\phi_{s}$ value between $0$ and $\pi/2$ will result in the generation of a partial interference pattern (see Appendix~\ref{Methods a} section for more details). Finally, the distinguishability can be assessed by blocking one of the paths in the unitary transformation stage using the variable electro-optical attenuators.

\begin{figure}[h]
\includegraphics[scale =0.34]{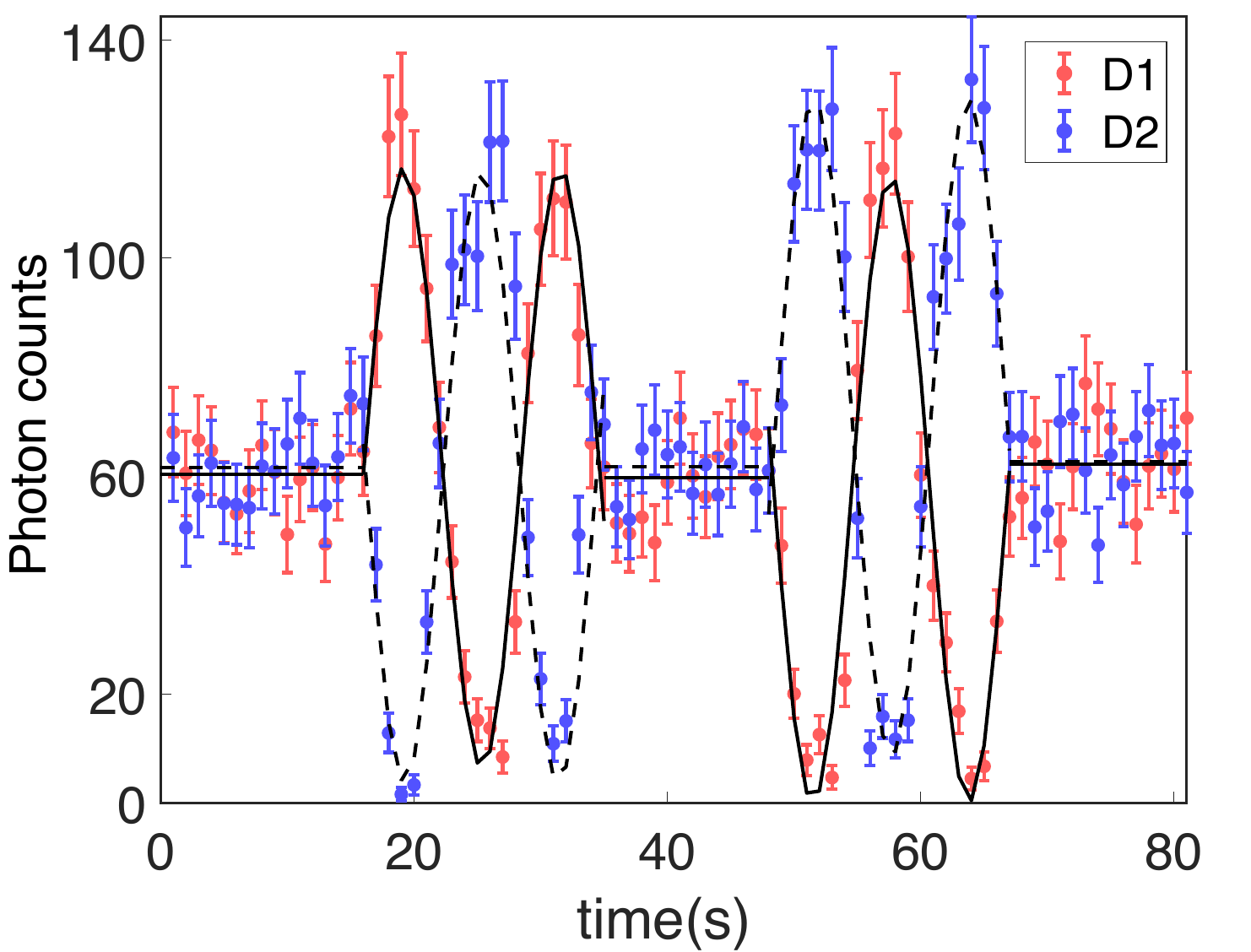}
\caption{Dynamic change between OAM wave and particle behavior. Recorded detections at D1 and D2 with a continuous triangle waveform applied to $\phi_x$ and $\phi_s$ periodically changed between $0$ (particle) or $\pi/2$ (wave). Error bars are the standard deviation considering Poissonian statistics of the single-photon detection process.}
\label{Fig3}
\end{figure}

An important feature of our experimental setup is the reconfigurability of the measurement from mirror to beam splitter mode due to the use of a fast electro-optical modulator controlling $\phi_s$. We demonstrate this dynamic reconfigurability by continuously modulating $\phi_x$ with a triangular waveform and switching $\phi_s$ from $0$ to $\pi/2$ every $18$ s approximately. These results can be observed in Fig. \ref{Fig3}, where we see the switching from particle to wave behavior owing to the state of the TBS.
\begin{figure*}[ht]
\centering
\includegraphics[width=18cm]{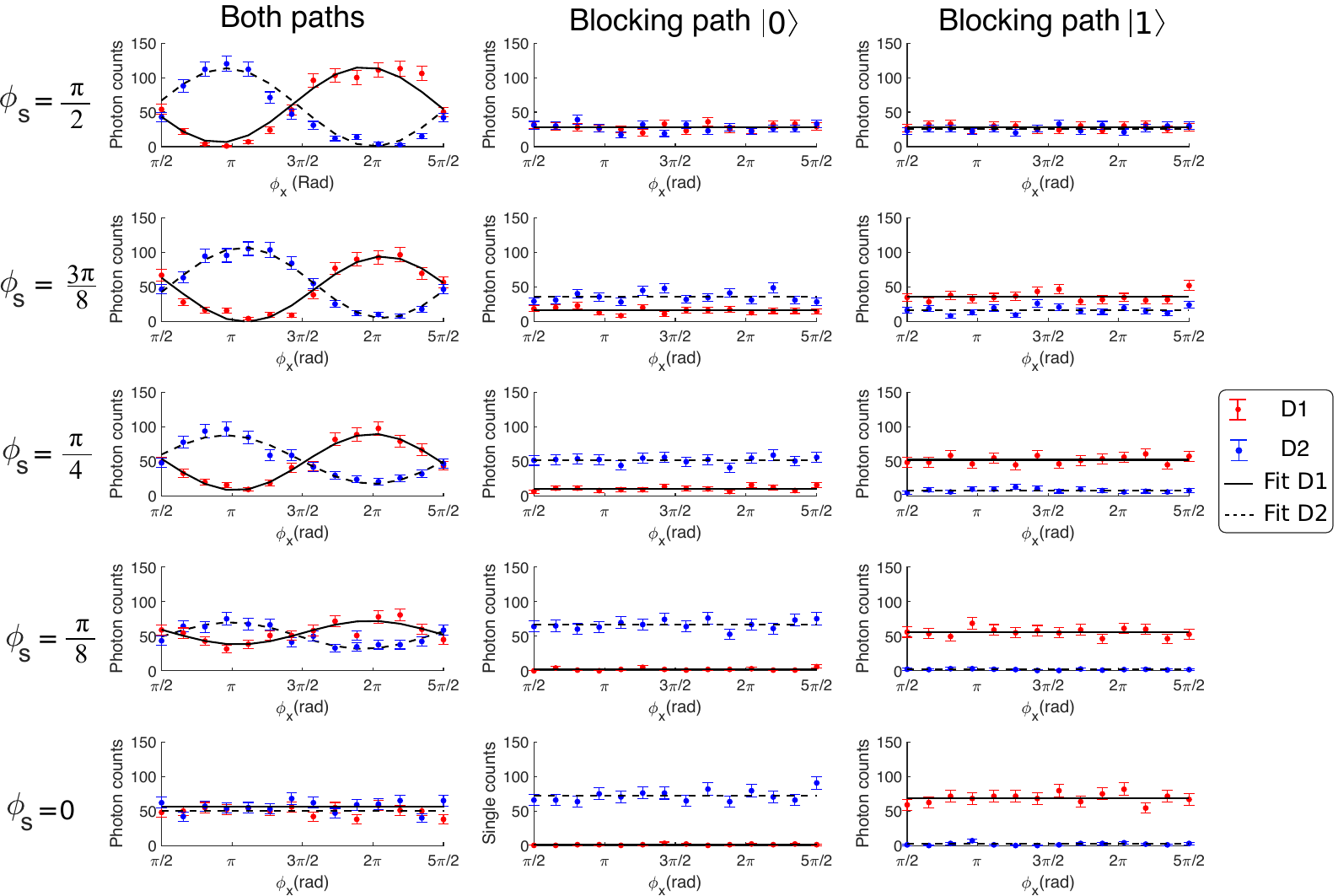}
\caption{Single counts for detectors $1$ and $2$ as a function of $\phi_{x}$, varying the phase shift $\phi_{s}$ applied in the Sagnac interferometer. In the first column, both arms of the interferometer are open, while the second and third columns show the single count when paths $0$ or $1$ are blocked, respectively. In each subplot, the error bars were calculated through error propagation taking into account the Poissonian statistics of the recorded individual counts. The integration time for each point is 0.8 s. Finally, the solid and dashed lines represent the best-fit curves obtained by minimizing the mean squared error between the experimental data and the fit.}
\label{Fig4}
\end{figure*}

To experimentally verify the equivalence between wave-particle duality and entropic uncertainty relation, we measure the interferometric visibility and distinguishability of the OAM state for different TBS configurations to evaluate the entropies, and we compare these measurements with the entropies being directly obtained from the probabilities (see Appendix~\ref{Methods a} section for the details). We adjust the phase parameter $\phi_s$ to nine different values during these measurements, while continuously varying $\phi_x$. We show
five different configurations of the TBS are shown in Fig. \ref{Fig4}, including the two extreme configurations $\phi_s=\pi/2$ (beam splitter mode) and $\phi_s=0$ (mirror mode). The first column is when both paths are open. The second (third) column is when the variable attenuator blocks the $|0\rangle$ ($|1\rangle$) path. Each plot shows the number of counts from D1 and D2 versus $\phi_x$. In the case where both paths are open, the maximum visibility is $96.7\pm2.3\%$. As expected, no interference pattern is observed when $\phi_{s}=0$. In addition, when one of the paths is blocked, no interference pattern is visible. However, a difference appears when the TBS is continuously switched from full beamsplitter to mirror mode when one of the paths is blocked. When $\phi_s =\pi/2$, the SI acts as a 50:50 BS, and the detectors both measure the same number of counts. As $\phi_{s}$ goes to $0$, the detection probability at one of the detectors gradually increases until all the counts are obtained in one detector at $\phi_s = 0$.

\begin{figure}[htp]
\includegraphics[scale =0.36]{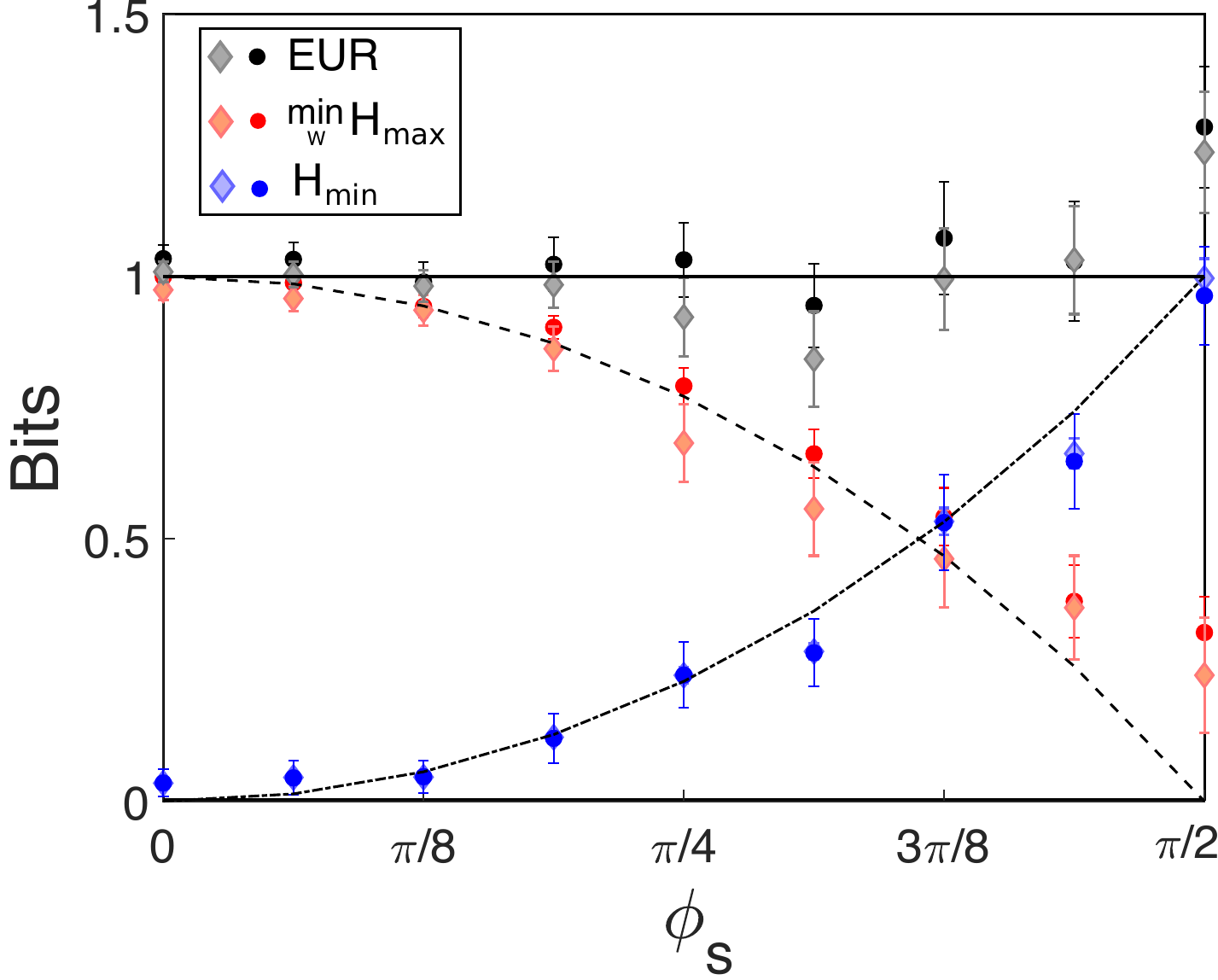}
\caption{Experimental equivalence between WPDR and EUR. The solid, dashed, and dot-dashed curves are respectively the theoretical values, as a function of the Sagnac phase $\phi_{s}$, of the EUR, minimized max-entropy related to the random variable $W$ (wave), and min-entropy related to the random variable $Z$ (particle). The black, blue, and red circles with their error bars are obtained, for each value of $\phi_{s}$, by measuring the input distinguishability $\mathcal{D}$ and the interferometric visibility $\mathcal{V}$, to evaluate the entropies via Eqs.~\ref{Eq6} and \ref{Eq7}. The black, blue, and red diamonds with their error bars are obtained by taking the measured probabilities related to each variable and applying the entropic definitions as given by Eqs.~\ref{eq:min-entropy-single} and \ref{eq:max-entropy-single}. Error bars were calculated using error propagation assuming Poissonian statistics for the recorded number of detection events.}
\label{Fig5}
\end{figure}

From the probabilities depicted in Fig.~\ref{Fig4}, the visibility (Eq.~\ref{Eq1}) and distinguishability (Eq.~\ref{Eq2}) are calculated for different TBS configurations, which are then used to calculate the min- and max-entropies according to Eqs.~\ref{Eq6} and \ref{Eq7}. From these two quantities, the entropic uncertainty relation (Eq.~\ref{EUR1}) is first obtained. To show the experimental equivalence, we also employ the final interferometric probabilities directly into the definitions of the unconditional min- and max-entropies respectively defined in Eqs.~\ref{eq:min-entropy-single} and \ref{eq:max-entropy-single}. 
Fig. \ref{Fig5} presents our main result: the WPDR based on input distinguishability and interferometric visibility is experimentally equivalent to the optimized EUR based on the unconditional min- and max-entropies. The discrepancy observed in the last measurement is due to the limited reach in the visibility of our setup, mainly given by modal crosstalk in the photonic lantern.

\section{Discussion}

Wave-particle duality is a well-known hallmark feature of quantum mechanics, and it limits the type of measurements one can do on a quantum system simultaneously. In Ref.~\cite{coles2014equivalence}, a theoretical framework was established to prove the equivalence between WPDRs and EURs. In this work, we experimentally verified the equivalence between the WPDR based on the input distinguishability $\mathcal{D}$ and interferometric visibility $\mathcal{V}$, with the optimized EUR based on the two random variables $Z$ (associated with particle behavior) and $W$ (associated with wave behavior). Our results corroborate the predictions and, once more, confirm Bohr's complementarity principle which states that individual quantum systems carry information that cannot be fully extracted from the same experimental arrangement. As shown in our main result (Fig.~\ref{Fig5}), regardless of the interferometer's configuration, there is always at least one bit of ignorance regarding the information associated with the random variables $Z$ and $W$ when measuring $\mathcal{D}$ and $\mathcal{V}$ directly.  Moreover, by comparing the two different ways of assessing the EUR, Fig. \ref{Fig5} shows that, up to error bars, both quantities are the same.
Beyond its foundational implications, these results are also highly relevant from a quantum information perspective as most quantum communication protocols rely on measurements with incompatible operators, thus providing a new perspective for practical applications based on the operational meaning of the min and max-entropies~\cite{coles2017entropic}.

We also have demonstrated a new technique to process OAM quantum states of light, based on few-mode fibers, a photonic lantern, and a Sagnac interferometer operating as a tunable beamsplitter. To the best of our knowledge, this is the first wave-particle duality measurements of OAM states in an all-fiber platform, opening up new possibilities to process OAM states with fast response times, due to the electro-optical modulators employed, a sharp contrast to the more usual bulk optics components such as q-plates and spatial light modulators. The fast response times of our setup can be used to expand this experiment to new applications such as delayed-choice quantum communication using OAM or path-encoded quantum states, giving new alternatives to novel protocols of quantum information.

\begin{appendix}

\section{EUR and WPDRs in interferometric settings}\label{Methods a}

Here, we define the relevant quantities and briefly explain the theoretical framework based on the operational meaning of the max and min-entropies and their connection with wave-particle duality~\cite{coles2014equivalence,coles2016entropic}. 
For a classical-quantum state $\rho_{\mathcal{AB}}=\sum_{a}p_a\ket{a}\bra{a}\otimes \rho_{\mathcal{B}}^a$ in a Hilbert space $\mathcal{H_A}^{(n)}\otimes\mathcal{H_B}^{(m)}$ such that $n$ and $m$ are the dimensions of each subspace, the min-entropy is defined as~\cite{konig2009operational}
\begin{equation}
    H_{\text{min}}(A|B):= -\log_2{p_{\text{guess}}(A|B)},
\end{equation}
where 
\begin{equation}
    p_{\text{guess}}(A|B):= \max_{\mathcal{M}_a}\sum_ap_a\text{Tr}[\mathbbm{M}_a\rho_{B}^a]
\end{equation}
is the probability of guessing $A$ correctly given the outcome of the optimal POVM measurement $\mathbbm{M}_a$ on system $B$. Note that, when the second subsystem is assumed to be trivial, the unconditional min-entropy for a given probability distribution $P=\{p_j\}$ reduces to
\begin{equation}
\label{eq:min-entropy-single}
     H_{\text{min}}(P):= -\log_2{\max_j p_j}.
\end{equation}

The max-entropy is defined as
\begin{equation}
    H_{\text{max}}(A|B):= \log_2{p_{\text{secr}}(A|B)},
\end{equation}
where
\begin{equation}
  p_{\text{secr}}(A|B):= \max_{\sigma_{\mathcal{B}}} F(\rho_{\mathcal{AB}}, \mathbbm{1}\otimes \sigma_{\mathcal{B}})  
\end{equation}
quantifies the secrecy of $A$ from $B$, as measured by the maximum possible fidelity of $\rho_{AB}$ to an uncorrelated state $\sigma_B$~\cite{coles2016entropic}. Moreover, the unconditional max-entropy is defined as
\begin{equation}
\label{eq:max-entropy-single}
    H_{\text{max}}(P):= 2\log_2{\sum_j  \sqrt{p_j} }.
\end{equation} 
For more details regarding applications such as how the min-entropy is employed in quantum key distribution to quantify how well the eavesdropper can guess the secret key, we refer the reader to Ref.~\cite{coles2017entropic}.

Equation \ref{eq: EUR} can be proved as a WPDR by examining the connection between the max-entropy and guessing probability by the upper bound introduced in Ref.~\cite{coles2016entropic}
\begin{equation}
    H_{\text{max}}(A|B)\leq \log_2{\left(1+\sqrt{(n-1)^2-(n p_{\text{guess}}(A|B)-1)^2}\right)},
\end{equation}
and noting that minimizing $\min_{\mathbbm{W}} H_{\text{max}}(W)$ means maximizing $p_{\text{guess}}(W)$ over all possible basis $\mathbbm{W}$. The equivalence of EUR and WPDR can then be demonstrated according to some particular interferometric setting. For instance, when we have a symmetric superposition of paths inside a multipath interferometer, Eq.~\ref{eq: EUR} translates to 
\begin{equation}
   \mathcal{D}^2(Z|B)+\mathcal{V}^2\leq 1.
\end{equation}
where
\begin{equation}
\label{eq:Vn}
    \mathcal{V}:=\frac{np_{\text{guess}}^{\max_{\{\phi_k\}}}(W)-1}{n-1},
\end{equation}
is the generalized notion of interferometric visibility related to the set of relative controllable phases $\{\phi_k\}$, that exhibits a trade-off with the generalized version of path distinguishability
\begin{equation}
\label{eq:Dn}
    \mathcal{D}(Z|B):= \frac{n p_{\text{guess}}(Z|B)-1}{n-1} .
\end{equation}

For binary interferometers and treating the second subsystem as trivial as stated in the EUR Eq.~\ref{EUR1}, Eqs.~\ref{eq:Vn} and~\ref{eq:Dn} can be experimentally obtained as respectively described in Eqs.~\ref{Eq1} and~\ref{Eq2} and depicted in Fig.~\ref{Fig1}. In the following, we detail their theoretical description using the states of the interferometer depicted in Fig. \ref{Fig2}. For this, we will use the well-known operators that allow mapping some optical components into their matrix representations. Note that, as depicted in Fig.~\ref{Fig2}, when the single photon enters the SI through BS2, its probability amplitudes recombine in BS2 itself after traveling the entire interferometer. This occurs both in the clockwise and counterclockwise directions, therefore, when producing interference at the output of the SI, a $\frac{\pi}{2}$-phase difference is considered. So, it follows that each one is modeled by 
\begin{equation}
\text{BS1} = \frac{1}{\sqrt{2}} \begin{pmatrix}
1 & i \\
i & 1
\end{pmatrix}, \quad
\text{BS2} = \frac{1}{\sqrt{2}} \begin{pmatrix}
i & -1 \\
-1 & i
\end{pmatrix}.
\end{equation}
Two quantum operators corresponding to both phase modulators are presented below
\begin{equation}
\text{PM1} = \begin{pmatrix}
1 & 0 \\
0 & e^{i\phi_x}
\end{pmatrix}, \quad
\text{PM2} = \begin{pmatrix}
1 & 0 \\
0 & e^{i\phi_S}
\end{pmatrix}.
\end{equation}

The interferometric output state can be written as
\begin{equation}
\begin{aligned}
        \ket{\psi}=&\frac{1}{2\sqrt{2}}[i((1+e^{i \phi_s})+e^{i \phi_x}(1-e^{i \phi_s}))\ket{D_1} \\
        &+((e^{i \phi_s}-1)-e^{i \phi_x}(1+e^{i \phi_s}))\ket{D_2}].
\end{aligned}
\end{equation}
The probabilities of detecting photons in $D_1$ and $D_2$ are
\begin{equation}
    p_1:=|\bra{D_1}\ket{\psi}|^2=\frac{2-\cos(\phi_x+\phi_s)+\cos(\phi_x-\phi_s)}{4},
    \label{Eq9}
\end{equation}
and
\begin{equation}
    p_2:=|\bra{D_2}\ket{\psi}|^2=\frac{2+\cos(\phi_x+\phi_s)-\cos(\phi_x-\phi_s)}{4}.
    \label{Eq10}
\end{equation}
From here, we can derive the expressions for the max-entropy by applying it to the state inside of the interferometer, it reads
\begin{equation}
\begin{aligned}
    \min_{\mathbbm{W}} H_{\text{max}}(W)&=\min_{\phi_x}\log_2{(1+\sqrt{1-\sin{\phi_s}^2\sin{\phi_x}^2})}\\
    &=\log_2{(1+\sqrt{1-\sin{\phi_s}^2})},
    \end{aligned}
\end{equation}
where $V=\sin{\phi_s}$.
On the other hand, distinguishability is obtained by either blocking the lower or the upper path in the first stage of our interferometer. For instance, by blocking the upper path of the state inside the interferometer, the output state reads
\begin{equation}
    \ket{\psi}=\frac{1}{2}[i(1+e^{i\phi_s})\ket{D_1}+(1-e^{i\phi_s})\ket{D_2}].
\end{equation}
We can observe that this equation is equivalent to that of a traditional SI when only one input is connected, which is an expected outcome when one of the paths of the interferometer is blocked, as depicted in Fig.\ref{Fig1}.d) and Fig.\ref{Fig1}.e). Furthermore, the wave function does not depend on $\phi_x$, as this relative phase arises only when both arms are connected. One can find the input distinguishability using the state inside the interferometer as
\begin{equation}
    H_{min}(Z)=-\log_2{\cos^2({\phi_s/2})}=-\log_2{(\frac{1+\cos{\phi_s}}{2})},
\end{equation}
such that $\mathcal{D}=\cos{\phi_s}$. The above relations for max and min-entropies as a function of $V$ and $D$ 
prove the equivalence of Eq.~\ref{EUR1} with 
\begin{equation}
    \mathcal{D}^2+\mathcal{V}^2\leq 1.
\end{equation}

\section{OAM state preparation} \label{Methods b}

To prepare the OAM state, it is necessary to create a helical mode. This is achieved using a computer-generated hologram displayed on a spatial light modulator (SLM), which functions as an optical device capable of converting a Gaussian mode into an OAM mode. For optimal operation of the SLM, a Gaussian beam must be propagated, and the laser's polarization must be set to horizontal. The Gaussian profile naturally comes from the single-mode fiber connected to the laser, while a polarizing beam splitter (PBS) in conjunction with a half-wave plate (HWP) in free space ensures horizontal polarization before the SLM (Fig. \ref{Fig2}). In addition, the beam is collimated using a $10\times$ objective, resulting in a beam waist of approximately $1250$ $\mu$m, which is sufficient to fully illuminate the forked hologram. The forked hologram consists of a helical phase profile superposed with a linear phase ramp to isolate the encoded field from the Gaussian mode, resulting in a diffraction grating that produces the OAM$_{+1}$ mode in the first diffraction order \cite{Alarcon2023}. To select the first order of the diffraction, a $4$f system is used. This system is composed of two lenses, L$1$ and L$2$, each with a focal length of $150$ mm. The Fourier transform of the reflected field is located in the focal plane of L$1$, where a pinhole is placed to pick out the first-order diffraction, which corresponds to the Fourier spectrum of the encoded field. Then, the second lens L$2$ performs an inverse Fourier transform to finally obtain the OAM$_{+1}$ mode. A $20\times$ objective is used to couple the OAM$_{+1}$ mode into a few-mode fiber (FMF), which is placed at the image plane of the $20\times$ objective lens. This procedure allows the generation of weak coherent states (WCSs) carrying OAM$_{+1}$, which are then propagated through the FMF.

\section{Single-detector scheme}\label{Methods c}

We employed one single-photon detection counting module to measure the detection probabilities at both outputs of our setup. In order to be able to simultaneously measure both outputs with a single detector we employ a time-multiplexing scheme \cite{Argillander_2022}. A relative optical fiber delay of 1250 ns is inserted at the D1 output in Fig. \ref{Fig1}, which is then connected to a fiber-optical polarising beam splitter (PBS), with the other input to the PBS connected to the D2 output. Manual polarisation controllers are used to optimize the transmission of both PBS inputs to the same common output, which is then connected to the single-photon detector. Depending on the arrival time (\textit{early} or \textit{late}), the processing electronics assign the corresponding output to the detection event. Therefore, for each created weak coherent state, the single-photon detector is gated twice, one for the \textit{early} and the other for the \textit{late} detection event. The delay between the two consecutive gates is sufficiently large to not have a significant afterpulse probability in the \textit{late} gate. This technique places an upper bound on the repetition rate of the experiment. This scheme is nevertheless also useful in many quantum key distribution schemes, as it prevents some side-channel attacks that depend on the physical asymmetries between two single-photon detectors \cite{Lo_timeshift}. 

\end{appendix}

\section*{Acknowledgements}
The authors wish to thank Gustaf \AA hlgren and Hilma Karlsson for earlier discussions. Financial support is acknowledged from Zenith Link\"{o}ping University and the Wallenberg Center for Quantum Technologies. Also, this work was supported by Fondo Nacional de Desarrollo Cientifico y Tecnológico (FONDECYT) (Grant no.3210359). M.P. acknowledges QuantERA, an ERA-Net cofund in Quantum Technologies, under the project eDICT. P.R.D acknowledges support from the NCN Poland, ChistEra-2023/05/Y/ST2/00005 under the project Modern Device Independent Cryptography (MoDIC).

\bibliography{common}

\begin{thebibliography}{54}%
\makeatletter
\providecommand \@ifxundefined [1]{%
 \@ifx{#1\undefined}
}%
\providecommand \@ifnum [1]{%
 \ifnum #1\expandafter \@firstoftwo
 \else \expandafter \@secondoftwo
 \fi
}%
\providecommand \@ifx [1]{%
 \ifx #1\expandafter \@firstoftwo
 \else \expandafter \@secondoftwo
 \fi
}%
\providecommand \natexlab [1]{#1}%
\providecommand \enquote  [1]{``#1''}%
\providecommand \bibnamefont  [1]{#1}%
\providecommand \bibfnamefont [1]{#1}%
\providecommand \citenamefont [1]{#1}%
\providecommand \href@noop [0]{\@secondoftwo}%
\providecommand \href [0]{\begingroup \@sanitize@url \@href}%
\providecommand \@href[1]{\@@startlink{#1}\@@href}%
\providecommand \@@href[1]{\endgroup#1\@@endlink}%
\providecommand \@sanitize@url [0]{\catcode `\\12\catcode `\$12\catcode `\&12\catcode `\#12\catcode `\^12\catcode `\_12\catcode `\%12\relax}%
\providecommand \@@startlink[1]{}%
\providecommand \@@endlink[0]{}%
\providecommand \url  [0]{\begingroup\@sanitize@url \@url }%
\providecommand \@url [1]{\endgroup\@href {#1}{\urlprefix }}%
\providecommand \urlprefix  [0]{URL }%
\providecommand \Eprint [0]{\href }%
\providecommand \doibase [0]{https://doi.org/}%
\providecommand \selectlanguage [0]{\@gobble}%
\providecommand \bibinfo  [0]{\@secondoftwo}%
\providecommand \bibfield  [0]{\@secondoftwo}%
\providecommand \translation [1]{[#1]}%
\providecommand \BibitemOpen [0]{}%
\providecommand \bibitemStop [0]{}%
\providecommand \bibitemNoStop [0]{.\EOS\space}%
\providecommand \EOS [0]{\spacefactor3000\relax}%
\providecommand \BibitemShut  [1]{\csname bibitem#1\endcsname}%
\let\auto@bib@innerbib\@empty
\bibitem [{\citenamefont {Dieguez}\ \emph {et~al.}(2022)\citenamefont {Dieguez}, \citenamefont {Guimar{\~a}es}, \citenamefont {Peterson}, \citenamefont {Angelo},\ and\ \citenamefont {Serra}}]{dieguez2022experimental}%
  \BibitemOpen
  \bibfield  {author} {\bibinfo {author} {\bibfnamefont {P.~R.}\ \bibnamefont {Dieguez}}, \bibinfo {author} {\bibfnamefont {J.~R.}\ \bibnamefont {Guimar{\~a}es}}, \bibinfo {author} {\bibfnamefont {J.~P.}\ \bibnamefont {Peterson}}, \bibinfo {author} {\bibfnamefont {R.~M.}\ \bibnamefont {Angelo}},\ and\ \bibinfo {author} {\bibfnamefont {R.~M.}\ \bibnamefont {Serra}},\ }\bibfield  {title} {\bibinfo {title} {Experimental assessment of physical realism in a quantum-controlled device},\ }\href {https://doi.org/10.1038/s42005-022-00828-z} {\bibfield  {journal} {\bibinfo  {journal} {Commun. Phys.}\ }\textbf {\bibinfo {volume} {5}},\ \bibinfo {pages} {82} (\bibinfo {year} {2022})}\BibitemShut {NoStop}%
\bibitem [{\citenamefont {Konig}\ \emph {et~al.}(2009)\citenamefont {Konig}, \citenamefont {Renner},\ and\ \citenamefont {Schaffner}}]{konig2009operational}%
  \BibitemOpen
  \bibfield  {author} {\bibinfo {author} {\bibfnamefont {R.}~\bibnamefont {Konig}}, \bibinfo {author} {\bibfnamefont {R.}~\bibnamefont {Renner}},\ and\ \bibinfo {author} {\bibfnamefont {C.}~\bibnamefont {Schaffner}},\ }\bibfield  {title} {\bibinfo {title} {The operational meaning of min-and max-entropy},\ }\href {https://doi.org/10.1109/TIT.2009.2025545} {\bibfield  {journal} {\bibinfo  {journal} {IEEE Transactions on Information theory}\ }\textbf {\bibinfo {volume} {55}},\ \bibinfo {pages} {4337} (\bibinfo {year} {2009})}\BibitemShut {NoStop}%
\bibitem [{\citenamefont {Coles}\ \emph {et~al.}(2017)\citenamefont {Coles}, \citenamefont {Berta}, \citenamefont {Tomamichel},\ and\ \citenamefont {Wehner}}]{coles2017entropic}%
  \BibitemOpen
  \bibfield  {author} {\bibinfo {author} {\bibfnamefont {P.~J.}\ \bibnamefont {Coles}}, \bibinfo {author} {\bibfnamefont {M.}~\bibnamefont {Berta}}, \bibinfo {author} {\bibfnamefont {M.}~\bibnamefont {Tomamichel}},\ and\ \bibinfo {author} {\bibfnamefont {S.}~\bibnamefont {Wehner}},\ }\bibfield  {title} {\bibinfo {title} {Entropic uncertainty relations and their applications},\ }\href {https://doi.org/10.1103/RevModPhys.89.015002} {\bibfield  {journal} {\bibinfo  {journal} {Reviews of Modern Physics}\ }\textbf {\bibinfo {volume} {89}},\ \bibinfo {pages} {015002} (\bibinfo {year} {2017})}\BibitemShut {NoStop}%
\bibitem [{\citenamefont {Coles}\ \emph {et~al.}(2014)\citenamefont {Coles}, \citenamefont {Kaniewski},\ and\ \citenamefont {Wehner}}]{coles2014equivalence}%
  \BibitemOpen
  \bibfield  {author} {\bibinfo {author} {\bibfnamefont {P.~J.}\ \bibnamefont {Coles}}, \bibinfo {author} {\bibfnamefont {J.}~\bibnamefont {Kaniewski}},\ and\ \bibinfo {author} {\bibfnamefont {S.}~\bibnamefont {Wehner}},\ }\bibfield  {title} {\bibinfo {title} {Equivalence of wave-particle duality to entropic uncertainty},\ }\href {https://doi.org/10.1038/ncomms6814} {\bibfield  {journal} {\bibinfo  {journal} {Nat. Commun.}\ }\textbf {\bibinfo {volume} {5}},\ \bibinfo {pages} {5814} (\bibinfo {year} {2014})}\BibitemShut {NoStop}%
\bibitem [{\citenamefont {Coles}(2016)}]{coles2016entropic}%
  \BibitemOpen
  \bibfield  {author} {\bibinfo {author} {\bibfnamefont {P.~J.}\ \bibnamefont {Coles}},\ }\bibfield  {title} {\bibinfo {title} {Entropic framework for wave-particle duality in multipath interferometers},\ }\href {https://doi.org/10.1103/PhysRevA.93.062111} {\bibfield  {journal} {\bibinfo  {journal} {Phys. Rev. A}\ }\textbf {\bibinfo {volume} {93}},\ \bibinfo {pages} {062111} (\bibinfo {year} {2016})}\BibitemShut {NoStop}%
\bibitem [{\citenamefont {Dieguez}\ and\ \citenamefont {Angelo}(2018)}]{dieguez2018information}%
  \BibitemOpen
  \bibfield  {author} {\bibinfo {author} {\bibfnamefont {P.~R.}\ \bibnamefont {Dieguez}}\ and\ \bibinfo {author} {\bibfnamefont {R.~M.}\ \bibnamefont {Angelo}},\ }\bibfield  {title} {\bibinfo {title} {Information-reality complementarity: {T}he role of measurements and quantum reference frames},\ }\href {https://doi.org/10.1103/PhysRevA.97.022107} {\bibfield  {journal} {\bibinfo  {journal} {Phys. Rev. A}\ }\textbf {\bibinfo {volume} {97}},\ \bibinfo {pages} {022107} (\bibinfo {year} {2018})}\BibitemShut {NoStop}%
\bibitem [{\citenamefont {Bohr}(1928)}]{bohr1928quantum}%
  \BibitemOpen
  \bibfield  {author} {\bibinfo {author} {\bibfnamefont {N.}~\bibnamefont {Bohr}},\ }\bibfield  {title} {\bibinfo {title} {The quantum postulate and the recent development of atomic theory},\ }\href {https://doi.org/10.1038/121580a0} {\bibfield  {journal} {\bibinfo  {journal} {Nature}\ }\textbf {\bibinfo {volume} {121}},\ \bibinfo {pages} {580} (\bibinfo {year} {1928})}\BibitemShut {NoStop}%
\bibitem [{\citenamefont {Bohr}(1935)}]{bohr1935can}%
  \BibitemOpen
  \bibfield  {author} {\bibinfo {author} {\bibfnamefont {N.}~\bibnamefont {Bohr}},\ }\bibfield  {title} {\bibinfo {title} {Can quantum-mechanical description of physical reality be considered complete?},\ }\href {https://doi.org/10.1103/PhysRev.48.696} {\bibfield  {journal} {\bibinfo  {journal} {Phys. Rev.}\ }\textbf {\bibinfo {volume} {48}},\ \bibinfo {pages} {696} (\bibinfo {year} {1935})}\BibitemShut {NoStop}%
\bibitem [{\citenamefont {Hameedi}\ \emph {et~al.}(2017)\citenamefont {Hameedi}, \citenamefont {Saha}, \citenamefont {Mironowicz}, \citenamefont {Paw{\l}owski},\ and\ \citenamefont {Bourennane}}]{hameedi2017complementarity}%
  \BibitemOpen
  \bibfield  {author} {\bibinfo {author} {\bibfnamefont {A.}~\bibnamefont {Hameedi}}, \bibinfo {author} {\bibfnamefont {D.}~\bibnamefont {Saha}}, \bibinfo {author} {\bibfnamefont {P.}~\bibnamefont {Mironowicz}}, \bibinfo {author} {\bibfnamefont {M.}~\bibnamefont {Paw{\l}owski}},\ and\ \bibinfo {author} {\bibfnamefont {M.}~\bibnamefont {Bourennane}},\ }\bibfield  {title} {\bibinfo {title} {Complementarity between entanglement-assisted and quantum distributed random access code},\ }\href {https://doi.org/10.1103/PhysRevA.95.052345} {\bibfield  {journal} {\bibinfo  {journal} {Physical Review A}\ }\textbf {\bibinfo {volume} {95}},\ \bibinfo {pages} {052345} (\bibinfo {year} {2017})}\BibitemShut {NoStop}%
\bibitem [{\citenamefont {Yadin}\ \emph {et~al.}(2021)\citenamefont {Yadin}, \citenamefont {Fadel},\ and\ \citenamefont {Gessner}}]{yadin2021metrological}%
  \BibitemOpen
  \bibfield  {author} {\bibinfo {author} {\bibfnamefont {B.}~\bibnamefont {Yadin}}, \bibinfo {author} {\bibfnamefont {M.}~\bibnamefont {Fadel}},\ and\ \bibinfo {author} {\bibfnamefont {M.}~\bibnamefont {Gessner}},\ }\bibfield  {title} {\bibinfo {title} {Metrological complementarity reveals the einstein-podolsky-rosen paradox},\ }\href {https://doi.org/10.1038/s41467-021-22353-3} {\bibfield  {journal} {\bibinfo  {journal} {Nature communications}\ }\textbf {\bibinfo {volume} {12}},\ \bibinfo {pages} {2410} (\bibinfo {year} {2021})}\BibitemShut {NoStop}%
\bibitem [{\citenamefont {Len}\ \emph {et~al.}(2022)\citenamefont {Len}, \citenamefont {Gefen}, \citenamefont {Retzker},\ and\ \citenamefont {Ko{\l}ody{\'n}ski}}]{len2022quantum}%
  \BibitemOpen
  \bibfield  {author} {\bibinfo {author} {\bibfnamefont {Y.~L.}\ \bibnamefont {Len}}, \bibinfo {author} {\bibfnamefont {T.}~\bibnamefont {Gefen}}, \bibinfo {author} {\bibfnamefont {A.}~\bibnamefont {Retzker}},\ and\ \bibinfo {author} {\bibfnamefont {J.}~\bibnamefont {Ko{\l}ody{\'n}ski}},\ }\bibfield  {title} {\bibinfo {title} {Quantum metrology with imperfect measurements},\ }\href {https://doi.org/10.1038/s41467-022-33563-8} {\bibfield  {journal} {\bibinfo  {journal} {Nature Communications}\ }\textbf {\bibinfo {volume} {13}},\ \bibinfo {pages} {6971} (\bibinfo {year} {2022})}\BibitemShut {NoStop}%
\bibitem [{\citenamefont {Scarani}\ \emph {et~al.}(2009)\citenamefont {Scarani}, \citenamefont {Bechmann-Pasquinucci}, \citenamefont {Cerf}, \citenamefont {Du{\v{s}}ek}, \citenamefont {L{\"u}tkenhaus},\ and\ \citenamefont {Peev}}]{scarani2009security}%
  \BibitemOpen
  \bibfield  {author} {\bibinfo {author} {\bibfnamefont {V.}~\bibnamefont {Scarani}}, \bibinfo {author} {\bibfnamefont {H.}~\bibnamefont {Bechmann-Pasquinucci}}, \bibinfo {author} {\bibfnamefont {N.~J.}\ \bibnamefont {Cerf}}, \bibinfo {author} {\bibfnamefont {M.}~\bibnamefont {Du{\v{s}}ek}}, \bibinfo {author} {\bibfnamefont {N.}~\bibnamefont {L{\"u}tkenhaus}},\ and\ \bibinfo {author} {\bibfnamefont {M.}~\bibnamefont {Peev}},\ }\bibfield  {title} {\bibinfo {title} {The security of practical quantum key distribution},\ }\href {https://doi.org/10.1103/RevModPhys.81.1301} {\bibfield  {journal} {\bibinfo  {journal} {Reviews of modern physics}\ }\textbf {\bibinfo {volume} {81}},\ \bibinfo {pages} {1301} (\bibinfo {year} {2009})}\BibitemShut {NoStop}%
\bibitem [{\citenamefont {Koashi}(2009)}]{koashi2009simple}%
  \BibitemOpen
  \bibfield  {author} {\bibinfo {author} {\bibfnamefont {M.}~\bibnamefont {Koashi}},\ }\bibfield  {title} {\bibinfo {title} {Simple security proof of quantum key distribution based on complementarity},\ }\href {https://doi.org/10.1088/1367-2630/11/4/045018} {\bibfield  {journal} {\bibinfo  {journal} {New Journal of Physics}\ }\textbf {\bibinfo {volume} {11}},\ \bibinfo {pages} {045018} (\bibinfo {year} {2009})}\BibitemShut {NoStop}%
\bibitem [{\citenamefont {Berta}\ \emph {et~al.}(2010)\citenamefont {Berta}, \citenamefont {Christandl}, \citenamefont {Colbeck}, \citenamefont {Renes},\ and\ \citenamefont {Renner}}]{berta2010uncertainty}%
  \BibitemOpen
  \bibfield  {author} {\bibinfo {author} {\bibfnamefont {M.}~\bibnamefont {Berta}}, \bibinfo {author} {\bibfnamefont {M.}~\bibnamefont {Christandl}}, \bibinfo {author} {\bibfnamefont {R.}~\bibnamefont {Colbeck}}, \bibinfo {author} {\bibfnamefont {J.~M.}\ \bibnamefont {Renes}},\ and\ \bibinfo {author} {\bibfnamefont {R.}~\bibnamefont {Renner}},\ }\bibfield  {title} {\bibinfo {title} {The uncertainty principle in the presence of quantum memory},\ }\href {https://doi.org/10.1038/nphys1734} {\bibfield  {journal} {\bibinfo  {journal} {Nature Physics}\ }\textbf {\bibinfo {volume} {6}},\ \bibinfo {pages} {659} (\bibinfo {year} {2010})}\BibitemShut {NoStop}%
\bibitem [{\citenamefont {Mizutani}\ \emph {et~al.}(2017)\citenamefont {Mizutani}, \citenamefont {Sasaki}, \citenamefont {Kato}, \citenamefont {Takeuchi},\ and\ \citenamefont {Tamaki}}]{mizutani2017information}%
  \BibitemOpen
  \bibfield  {author} {\bibinfo {author} {\bibfnamefont {A.}~\bibnamefont {Mizutani}}, \bibinfo {author} {\bibfnamefont {T.}~\bibnamefont {Sasaki}}, \bibinfo {author} {\bibfnamefont {G.}~\bibnamefont {Kato}}, \bibinfo {author} {\bibfnamefont {Y.}~\bibnamefont {Takeuchi}},\ and\ \bibinfo {author} {\bibfnamefont {K.}~\bibnamefont {Tamaki}},\ }\bibfield  {title} {\bibinfo {title} {Information-theoretic security proof of differential-phase-shift quantum key distribution protocol based on complementarity},\ }\href {https://doi.org/10.1088/2058-9565/aa8705} {\bibfield  {journal} {\bibinfo  {journal} {Quantum Science and Technology}\ }\textbf {\bibinfo {volume} {3}},\ \bibinfo {pages} {014003} (\bibinfo {year} {2017})}\BibitemShut {NoStop}%
\bibitem [{\citenamefont {Zhang}\ \emph {et~al.}(2023)\citenamefont {Zhang}, \citenamefont {Zeng}, \citenamefont {Ye}, \citenamefont {Lo},\ and\ \citenamefont {Ma}}]{zhang2023quantum}%
  \BibitemOpen
  \bibfield  {author} {\bibinfo {author} {\bibfnamefont {X.}~\bibnamefont {Zhang}}, \bibinfo {author} {\bibfnamefont {P.}~\bibnamefont {Zeng}}, \bibinfo {author} {\bibfnamefont {T.}~\bibnamefont {Ye}}, \bibinfo {author} {\bibfnamefont {H.-K.}\ \bibnamefont {Lo}},\ and\ \bibinfo {author} {\bibfnamefont {X.}~\bibnamefont {Ma}},\ }\bibfield  {title} {\bibinfo {title} {Quantum complementarity approach to device-independent security},\ }\href {https://doi.org/10.1103/PhysRevLett.131.140801} {\bibfield  {journal} {\bibinfo  {journal} {Physical Review Letters}\ }\textbf {\bibinfo {volume} {131}},\ \bibinfo {pages} {140801} (\bibinfo {year} {2023})}\BibitemShut {NoStop}%
\bibitem [{\citenamefont {Pal}\ \emph {et~al.}(2020)\citenamefont {Pal}, \citenamefont {Saryal}, \citenamefont {Segal}, \citenamefont {Mahesh},\ and\ \citenamefont {Agarwalla}}]{pal2020experimental}%
  \BibitemOpen
  \bibfield  {author} {\bibinfo {author} {\bibfnamefont {S.}~\bibnamefont {Pal}}, \bibinfo {author} {\bibfnamefont {S.}~\bibnamefont {Saryal}}, \bibinfo {author} {\bibfnamefont {D.}~\bibnamefont {Segal}}, \bibinfo {author} {\bibfnamefont {T.}~\bibnamefont {Mahesh}},\ and\ \bibinfo {author} {\bibfnamefont {B.~K.}\ \bibnamefont {Agarwalla}},\ }\bibfield  {title} {\bibinfo {title} {Experimental study of the thermodynamic uncertainty relation},\ }\href {https://doi.org/10.1103/PhysRevResearch.2.022044} {\bibfield  {journal} {\bibinfo  {journal} {Physical Review Research}\ }\textbf {\bibinfo {volume} {2}},\ \bibinfo {pages} {022044} (\bibinfo {year} {2020})}\BibitemShut {NoStop}%
\bibitem [{\citenamefont {Koyuk}\ and\ \citenamefont {Seifert}(2020)}]{koyuk2020thermodynamic}%
  \BibitemOpen
  \bibfield  {author} {\bibinfo {author} {\bibfnamefont {T.}~\bibnamefont {Koyuk}}\ and\ \bibinfo {author} {\bibfnamefont {U.}~\bibnamefont {Seifert}},\ }\bibfield  {title} {\bibinfo {title} {Thermodynamic uncertainty relation for time-dependent driving},\ }\href {https://doi.org/10.1103/PhysRevLett.125.260604} {\bibfield  {journal} {\bibinfo  {journal} {Physical Review Letters}\ }\textbf {\bibinfo {volume} {125}},\ \bibinfo {pages} {260604} (\bibinfo {year} {2020})}\BibitemShut {NoStop}%
\bibitem [{\citenamefont {Horowitz}\ and\ \citenamefont {Gingrich}(2020)}]{horowitz2020thermodynamic}%
  \BibitemOpen
  \bibfield  {author} {\bibinfo {author} {\bibfnamefont {J.~M.}\ \bibnamefont {Horowitz}}\ and\ \bibinfo {author} {\bibfnamefont {T.~R.}\ \bibnamefont {Gingrich}},\ }\bibfield  {title} {\bibinfo {title} {Thermodynamic uncertainty relations constrain non-equilibrium fluctuations},\ }\href {https://doi.org/10.1038/s41567-020-0853-5} {\bibfield  {journal} {\bibinfo  {journal} {Nature Physics}\ }\textbf {\bibinfo {volume} {16}},\ \bibinfo {pages} {15} (\bibinfo {year} {2020})}\BibitemShut {NoStop}%
\bibitem [{\citenamefont {Falasco}\ \emph {et~al.}(2020)\citenamefont {Falasco}, \citenamefont {Esposito},\ and\ \citenamefont {Delvenne}}]{falasco2020unifying}%
  \BibitemOpen
  \bibfield  {author} {\bibinfo {author} {\bibfnamefont {G.}~\bibnamefont {Falasco}}, \bibinfo {author} {\bibfnamefont {M.}~\bibnamefont {Esposito}},\ and\ \bibinfo {author} {\bibfnamefont {J.-C.}\ \bibnamefont {Delvenne}},\ }\bibfield  {title} {\bibinfo {title} {Unifying thermodynamic uncertainty relations},\ }\href {https://doi.org/10.1088/1367-2630/ab8679} {\bibfield  {journal} {\bibinfo  {journal} {New Journal of Physics}\ }\textbf {\bibinfo {volume} {22}},\ \bibinfo {pages} {053046} (\bibinfo {year} {2020})}\BibitemShut {NoStop}%
\bibitem [{\citenamefont {Vieira}\ \emph {et~al.}(2023)\citenamefont {Vieira}, \citenamefont {de~Oliveira}, \citenamefont {Santos}, \citenamefont {Dieguez},\ and\ \citenamefont {Serra}}]{vieira2023exploring}%
  \BibitemOpen
  \bibfield  {author} {\bibinfo {author} {\bibfnamefont {C.~H.}\ \bibnamefont {Vieira}}, \bibinfo {author} {\bibfnamefont {J.~L.}\ \bibnamefont {de~Oliveira}}, \bibinfo {author} {\bibfnamefont {J.~F.}\ \bibnamefont {Santos}}, \bibinfo {author} {\bibfnamefont {P.~R.}\ \bibnamefont {Dieguez}},\ and\ \bibinfo {author} {\bibfnamefont {R.~M.}\ \bibnamefont {Serra}},\ }\bibfield  {title} {\bibinfo {title} {Exploring quantum thermodynamics with nmr},\ }\href {https://doi.org/10.1016/j.jmro.2023.100105} {\bibfield  {journal} {\bibinfo  {journal} {Journal of Magnetic Resonance Open}\ }\textbf {\bibinfo {volume} {16}},\ \bibinfo {pages} {100105} (\bibinfo {year} {2023})}\BibitemShut {NoStop}%
\bibitem [{\citenamefont {Wheeler}\ and\ \citenamefont {Zurek}(2014)}]{Wheeler}%
  \BibitemOpen
  \bibfield  {author} {\bibinfo {author} {\bibfnamefont {J.~A.}\ \bibnamefont {Wheeler}}\ and\ \bibinfo {author} {\bibfnamefont {W.~H.}\ \bibnamefont {Zurek}},\ }\href {https://doi.org/10.1515/9781400854554} {\emph {\bibinfo {title} {Quantum theory and measurement}}},\ Vol.~\bibinfo {volume} {15}\ (\bibinfo  {publisher} {Princeton University Press},\ \bibinfo {year} {2014})\BibitemShut {NoStop}%
\bibitem [{\citenamefont {Jacques}\ \emph {et~al.}(2007)\citenamefont {Jacques}, \citenamefont {Wu}, \citenamefont {Grosshans}, \citenamefont {Treussart}, \citenamefont {Grangier}, \citenamefont {Aspect},\ and\ \citenamefont {Roch}}]{jackes2007experimental}%
  \BibitemOpen
  \bibfield  {author} {\bibinfo {author} {\bibfnamefont {V.}~\bibnamefont {Jacques}}, \bibinfo {author} {\bibfnamefont {E.}~\bibnamefont {Wu}}, \bibinfo {author} {\bibfnamefont {F.}~\bibnamefont {Grosshans}}, \bibinfo {author} {\bibfnamefont {F.}~\bibnamefont {Treussart}}, \bibinfo {author} {\bibfnamefont {P.}~\bibnamefont {Grangier}}, \bibinfo {author} {\bibfnamefont {A.}~\bibnamefont {Aspect}},\ and\ \bibinfo {author} {\bibfnamefont {J.-F.}\ \bibnamefont {Roch}},\ }\bibfield  {title} {\bibinfo {title} {Experimental realization of {W}heeler's delayed-choice gedanken experiment},\ }\href {https://doi.org/10.1126/science.1136303} {\bibfield  {journal} {\bibinfo  {journal} {Science}\ }\textbf {\bibinfo {volume} {315}},\ \bibinfo {pages} {966} (\bibinfo {year} {2007})}\BibitemShut {NoStop}%
\bibitem [{\citenamefont {Peres}(2000)}]{Peres00}%
  \BibitemOpen
  \bibfield  {author} {\bibinfo {author} {\bibfnamefont {A.}~\bibnamefont {Peres}},\ }\bibfield  {title} {\bibinfo {title} {Delayed choice for entanglement swapping},\ }\href@noop {} {\bibfield  {journal} {\bibinfo  {journal} {Journal of Modern Optics}\ }\textbf {\bibinfo {volume} {47}},\ \bibinfo {pages} {139} (\bibinfo {year} {2000})}\BibitemShut {NoStop}%
\bibitem [{\citenamefont {Jennewein}\ \emph {et~al.}(2005)\citenamefont {Jennewein}, \citenamefont {Brukner}, \citenamefont {Aspelmeyer},\ and\ \citenamefont {Zeilinger}}]{Jennewein05}%
  \BibitemOpen
  \bibfield  {author} {\bibinfo {author} {\bibfnamefont {T.}~\bibnamefont {Jennewein}}, \bibinfo {author} {\bibfnamefont {{\v{C}}.}~\bibnamefont {Brukner}}, \bibinfo {author} {\bibfnamefont {M.}~\bibnamefont {Aspelmeyer}},\ and\ \bibinfo {author} {\bibfnamefont {A.}~\bibnamefont {Zeilinger}},\ }\bibfield  {title} {\bibinfo {title} {Experimental proposal of switched ``delayed-choice'' for entanglement swapping},\ }\href@noop {} {\bibfield  {journal} {\bibinfo  {journal} {International Journal of Quantum Information}\ }\textbf {\bibinfo {volume} {3}},\ \bibinfo {pages} {73} (\bibinfo {year} {2005})}\BibitemShut {NoStop}%
\bibitem [{\citenamefont {Ma}\ \emph {et~al.}(2012)\citenamefont {Ma}, \citenamefont {Zotter}, \citenamefont {Kofler}, \citenamefont {Ursin}, \citenamefont {Jennewein}, \citenamefont {Brukner},\ and\ \citenamefont {Zeilinger}}]{Ma12}%
  \BibitemOpen
  \bibfield  {author} {\bibinfo {author} {\bibfnamefont {X.-s.}\ \bibnamefont {Ma}}, \bibinfo {author} {\bibfnamefont {S.}~\bibnamefont {Zotter}}, \bibinfo {author} {\bibfnamefont {J.}~\bibnamefont {Kofler}}, \bibinfo {author} {\bibfnamefont {R.}~\bibnamefont {Ursin}}, \bibinfo {author} {\bibfnamefont {T.}~\bibnamefont {Jennewein}}, \bibinfo {author} {\bibfnamefont {{\v{C}}.}~\bibnamefont {Brukner}},\ and\ \bibinfo {author} {\bibfnamefont {A.}~\bibnamefont {Zeilinger}},\ }\bibfield  {title} {\bibinfo {title} {Experimental delayed-choice entanglement swapping},\ }\href@noop {} {\bibfield  {journal} {\bibinfo  {journal} {Nature Physics}\ }\textbf {\bibinfo {volume} {8}},\ \bibinfo {pages} {479} (\bibinfo {year} {2012})}\BibitemShut {NoStop}%
\bibitem [{\citenamefont {Peres}\ and\ \citenamefont {Terno}(2004)}]{Terno}%
  \BibitemOpen
  \bibfield  {author} {\bibinfo {author} {\bibfnamefont {A.}~\bibnamefont {Peres}}\ and\ \bibinfo {author} {\bibfnamefont {D.~R.}\ \bibnamefont {Terno}},\ }\bibfield  {title} {\bibinfo {title} {Quantum information and relativity theory},\ }\href {https://doi.org/10.1103/RevModPhys.76.93} {\bibfield  {journal} {\bibinfo  {journal} {Rev. Mod. Phys.}\ }\textbf {\bibinfo {volume} {76}},\ \bibinfo {pages} {93} (\bibinfo {year} {2004})}\BibitemShut {NoStop}%
\bibitem [{\citenamefont {Auccaise}\ \emph {et~al.}(2012)\citenamefont {Auccaise}, \citenamefont {Serra}, \citenamefont {Filgueiras}, \citenamefont {Sarthour}, \citenamefont {Oliveira},\ and\ \citenamefont {C\'eleri}}]{auccaise2012experimental}%
  \BibitemOpen
  \bibfield  {author} {\bibinfo {author} {\bibfnamefont {R.}~\bibnamefont {Auccaise}}, \bibinfo {author} {\bibfnamefont {R.~M.}\ \bibnamefont {Serra}}, \bibinfo {author} {\bibfnamefont {J.~G.}\ \bibnamefont {Filgueiras}}, \bibinfo {author} {\bibfnamefont {R.~S.}\ \bibnamefont {Sarthour}}, \bibinfo {author} {\bibfnamefont {I.~S.}\ \bibnamefont {Oliveira}},\ and\ \bibinfo {author} {\bibfnamefont {L.~C.}\ \bibnamefont {C\'eleri}},\ }\bibfield  {title} {\bibinfo {title} {Experimental analysis of the quantum complementarity principle},\ }\href {https://doi.org/10.1103/PhysRevA.85.032121} {\bibfield  {journal} {\bibinfo  {journal} {Phys. Rev. A}\ }\textbf {\bibinfo {volume} {85}},\ \bibinfo {pages} {032121} (\bibinfo {year} {2012})}\BibitemShut {NoStop}%
\bibitem [{\citenamefont {Vedovato}\ \emph {et~al.}(2017)\citenamefont {Vedovato}, \citenamefont {Agnesi}, \citenamefont {Schiavon}, \citenamefont {Dequal}, \citenamefont {Calderaro}, \citenamefont {Tomasin}, \citenamefont {Marangon}, \citenamefont {Stanco}, \citenamefont {Luceri}, \citenamefont {Bianco} \emph {et~al.}}]{vedovato2020extending}%
  \BibitemOpen
  \bibfield  {author} {\bibinfo {author} {\bibfnamefont {F.}~\bibnamefont {Vedovato}}, \bibinfo {author} {\bibfnamefont {C.}~\bibnamefont {Agnesi}}, \bibinfo {author} {\bibfnamefont {M.}~\bibnamefont {Schiavon}}, \bibinfo {author} {\bibfnamefont {D.}~\bibnamefont {Dequal}}, \bibinfo {author} {\bibfnamefont {L.}~\bibnamefont {Calderaro}}, \bibinfo {author} {\bibfnamefont {M.}~\bibnamefont {Tomasin}}, \bibinfo {author} {\bibfnamefont {D.~G.}\ \bibnamefont {Marangon}}, \bibinfo {author} {\bibfnamefont {A.}~\bibnamefont {Stanco}}, \bibinfo {author} {\bibfnamefont {V.}~\bibnamefont {Luceri}}, \bibinfo {author} {\bibfnamefont {G.}~\bibnamefont {Bianco}}, \emph {et~al.},\ }\bibfield  {title} {\bibinfo {title} {Extending {W}heeler's delayed-choice experiment to space},\ }\href {https://doi.org/10.1126/sciadv.1701180} {\bibfield  {journal} {\bibinfo  {journal} {Sci. Adv.}\ }\textbf {\bibinfo {volume} {3}},\ \bibinfo {pages} {e1701180} (\bibinfo {year} {2017})}\BibitemShut {NoStop}%
\bibitem [{\citenamefont {Catani}\ \emph {et~al.}(2021)\citenamefont {Catani}, \citenamefont {Leifer}, \citenamefont {Schmid},\ and\ \citenamefont {Spekkens}}]{catani2021interference}%
  \BibitemOpen
  \bibfield  {author} {\bibinfo {author} {\bibfnamefont {L.}~\bibnamefont {Catani}}, \bibinfo {author} {\bibfnamefont {M.}~\bibnamefont {Leifer}}, \bibinfo {author} {\bibfnamefont {D.}~\bibnamefont {Schmid}},\ and\ \bibinfo {author} {\bibfnamefont {R.~W.}\ \bibnamefont {Spekkens}},\ }\bibfield  {title} {\bibinfo {title} {Why interference phenomena do not capture the essence of quantum theory},\ }\href {https://arxiv.org/abs/2111.13727} {\bibfield  {journal} {\bibinfo  {journal} {arXiv:2111.13727}\ } (\bibinfo {year} {2021})}\BibitemShut {NoStop}%
\bibitem [{\citenamefont {Catani}\ \emph {et~al.}(2022)\citenamefont {Catani}, \citenamefont {Leifer}, \citenamefont {Scala}, \citenamefont {Schmid},\ and\ \citenamefont {Spekkens}}]{catani2022aspects}%
  \BibitemOpen
  \bibfield  {author} {\bibinfo {author} {\bibfnamefont {L.}~\bibnamefont {Catani}}, \bibinfo {author} {\bibfnamefont {M.}~\bibnamefont {Leifer}}, \bibinfo {author} {\bibfnamefont {G.}~\bibnamefont {Scala}}, \bibinfo {author} {\bibfnamefont {D.}~\bibnamefont {Schmid}},\ and\ \bibinfo {author} {\bibfnamefont {R.~W.}\ \bibnamefont {Spekkens}},\ }\bibfield  {title} {\bibinfo {title} {What aspects of the phenomenology of interference witness nonclassicality?},\ }\href {https://arxiv.org/abs/2211.09850} {\bibfield  {journal} {\bibinfo  {journal} {arXiv:2211.09850}\ } (\bibinfo {year} {2022})}\BibitemShut {NoStop}%
\bibitem [{\citenamefont {Chrysosthemos}\ \emph {et~al.}(2023)\citenamefont {Chrysosthemos}, \citenamefont {Basso},\ and\ \citenamefont {Maziero}}]{chrysosthemos2023updating}%
  \BibitemOpen
  \bibfield  {author} {\bibinfo {author} {\bibfnamefont {D.~S.}\ \bibnamefont {Chrysosthemos}}, \bibinfo {author} {\bibfnamefont {M.~L.}\ \bibnamefont {Basso}},\ and\ \bibinfo {author} {\bibfnamefont {J.}~\bibnamefont {Maziero}},\ }\bibfield  {title} {\bibinfo {title} {Updating bohr's complementarity principle},\ }\href@noop {} {\bibfield  {journal} {\bibinfo  {journal} {arXiv preprint arXiv:2312.02743}\ } (\bibinfo {year} {2023})}\BibitemShut {NoStop}%
\bibitem [{\citenamefont {Allen}\ \emph {et~al.}(1992)\citenamefont {Allen}, \citenamefont {Beijersbergen}, \citenamefont {Spreeuw},\ and\ \citenamefont {Woerdman}}]{Allen:92}%
  \BibitemOpen
  \bibfield  {author} {\bibinfo {author} {\bibfnamefont {L.}~\bibnamefont {Allen}}, \bibinfo {author} {\bibfnamefont {M.~W.}\ \bibnamefont {Beijersbergen}}, \bibinfo {author} {\bibfnamefont {R.~J.~C.}\ \bibnamefont {Spreeuw}},\ and\ \bibinfo {author} {\bibfnamefont {J.~P.}\ \bibnamefont {Woerdman}},\ }\bibfield  {title} {\bibinfo {title} {Orbital angular momentum of light and the transformation of laguerre-gaussian laser modes},\ }\href {https://doi.org/10.1103/PhysRevA.45.8185} {\bibfield  {journal} {\bibinfo  {journal} {Phys. Rev. A}\ }\textbf {\bibinfo {volume} {45}},\ \bibinfo {pages} {8185} (\bibinfo {year} {1992})}\BibitemShut {NoStop}%
\bibitem [{\citenamefont {Shen}\ \emph {et~al.}(2019)\citenamefont {Shen}, \citenamefont {Wang}, \citenamefont {Xie}, \citenamefont {Min}, \citenamefont {Fu}, \citenamefont {Liu}, \citenamefont {Gong},\ and\ \citenamefont {Yuan}}]{Shen:19}%
  \BibitemOpen
  \bibfield  {author} {\bibinfo {author} {\bibfnamefont {Y.}~\bibnamefont {Shen}}, \bibinfo {author} {\bibfnamefont {X.}~\bibnamefont {Wang}}, \bibinfo {author} {\bibfnamefont {Z.}~\bibnamefont {Xie}}, \bibinfo {author} {\bibfnamefont {C.}~\bibnamefont {Min}}, \bibinfo {author} {\bibfnamefont {X.}~\bibnamefont {Fu}}, \bibinfo {author} {\bibfnamefont {Q.}~\bibnamefont {Liu}}, \bibinfo {author} {\bibfnamefont {M.}~\bibnamefont {Gong}},\ and\ \bibinfo {author} {\bibfnamefont {X.}~\bibnamefont {Yuan}},\ }\bibfield  {title} {\bibinfo {title} {Optical vortices 30 years on: Oam manipulation from topological charge to multiple singularities},\ }\href {https://doi.org/10.1038/s41377-019-0194-2} {\bibfield  {journal} {\bibinfo  {journal} {Light: Science {\&} Applications}\ }\textbf {\bibinfo {volume} {8}},\ \bibinfo {pages} {90} (\bibinfo {year} {2019})}\BibitemShut {NoStop}%
\bibitem [{\citenamefont {Erhard}\ \emph {et~al.}(2018)\citenamefont {Erhard}, \citenamefont {Fickler}, \citenamefont {Krenn},\ and\ \citenamefont {Zeilinger}}]{Erhard:18}%
  \BibitemOpen
  \bibfield  {author} {\bibinfo {author} {\bibfnamefont {M.}~\bibnamefont {Erhard}}, \bibinfo {author} {\bibfnamefont {R.}~\bibnamefont {Fickler}}, \bibinfo {author} {\bibfnamefont {M.}~\bibnamefont {Krenn}},\ and\ \bibinfo {author} {\bibfnamefont {A.}~\bibnamefont {Zeilinger}},\ }\bibfield  {title} {\bibinfo {title} {Twisted photons: new quantum perspectives in high dimensions},\ }\href {https://doi.org/10.1038/lsa.2017.146} {\bibfield  {journal} {\bibinfo  {journal} {Light: Science {\&} Applications}\ }\textbf {\bibinfo {volume} {7}},\ \bibinfo {pages} {17146} (\bibinfo {year} {2018})}\BibitemShut {NoStop}%
\bibitem [{\citenamefont {Leach}\ \emph {et~al.}(2002)\citenamefont {Leach}, \citenamefont {Padgett}, \citenamefont {Barnett}, \citenamefont {Franke-Arnold},\ and\ \citenamefont {Courtial}}]{Leach:02}%
  \BibitemOpen
  \bibfield  {author} {\bibinfo {author} {\bibfnamefont {J.}~\bibnamefont {Leach}}, \bibinfo {author} {\bibfnamefont {M.~J.}\ \bibnamefont {Padgett}}, \bibinfo {author} {\bibfnamefont {S.~M.}\ \bibnamefont {Barnett}}, \bibinfo {author} {\bibfnamefont {S.}~\bibnamefont {Franke-Arnold}},\ and\ \bibinfo {author} {\bibfnamefont {J.}~\bibnamefont {Courtial}},\ }\bibfield  {title} {\bibinfo {title} {Measuring the orbital angular momentum of a single photon},\ }\href {https://doi.org/10.1103/PhysRevLett.88.257901} {\bibfield  {journal} {\bibinfo  {journal} {Phys. Rev. Lett.}\ }\textbf {\bibinfo {volume} {88}},\ \bibinfo {pages} {257901} (\bibinfo {year} {2002})}\BibitemShut {NoStop}%
\bibitem [{\citenamefont {Karimi}\ \emph {et~al.}(2009)\citenamefont {Karimi}, \citenamefont {Piccirillo}, \citenamefont {Nagali}, \citenamefont {Marrucci},\ and\ \citenamefont {Santamato}}]{Karimi:09}%
  \BibitemOpen
  \bibfield  {author} {\bibinfo {author} {\bibfnamefont {E.}~\bibnamefont {Karimi}}, \bibinfo {author} {\bibfnamefont {B.}~\bibnamefont {Piccirillo}}, \bibinfo {author} {\bibfnamefont {E.}~\bibnamefont {Nagali}}, \bibinfo {author} {\bibfnamefont {L.}~\bibnamefont {Marrucci}},\ and\ \bibinfo {author} {\bibfnamefont {E.}~\bibnamefont {Santamato}},\ }\bibfield  {title} {\bibinfo {title} {{Efficient generation and sorting of orbital angular momentum eigenmodes of light by thermally tuned q-plates}},\ }\href {https://doi.org/10.1063/1.3154549} {\bibfield  {journal} {\bibinfo  {journal} {Appl. Phys. Lett.}\ }\textbf {\bibinfo {volume} {94}},\ \bibinfo {pages} {231124} (\bibinfo {year} {2009})}\BibitemShut {NoStop}%
\bibitem [{\citenamefont {Wang}\ \emph {et~al.}(2012)\citenamefont {Wang}, \citenamefont {Yang}, \citenamefont {Fazal}, \citenamefont {Ahmed}, \citenamefont {Yan}, \citenamefont {Huang}, \citenamefont {Ren}, \citenamefont {Yue}, \citenamefont {Dolinar}, \citenamefont {Tur},\ and\ \citenamefont {Willner}}]{Wang:12}%
  \BibitemOpen
  \bibfield  {author} {\bibinfo {author} {\bibfnamefont {J.}~\bibnamefont {Wang}}, \bibinfo {author} {\bibfnamefont {J.-Y.}\ \bibnamefont {Yang}}, \bibinfo {author} {\bibfnamefont {I.~M.}\ \bibnamefont {Fazal}}, \bibinfo {author} {\bibfnamefont {N.}~\bibnamefont {Ahmed}}, \bibinfo {author} {\bibfnamefont {Y.}~\bibnamefont {Yan}}, \bibinfo {author} {\bibfnamefont {H.}~\bibnamefont {Huang}}, \bibinfo {author} {\bibfnamefont {Y.}~\bibnamefont {Ren}}, \bibinfo {author} {\bibfnamefont {Y.}~\bibnamefont {Yue}}, \bibinfo {author} {\bibfnamefont {S.}~\bibnamefont {Dolinar}}, \bibinfo {author} {\bibfnamefont {M.}~\bibnamefont {Tur}},\ and\ \bibinfo {author} {\bibfnamefont {A.~E.}\ \bibnamefont {Willner}},\ }\bibfield  {title} {\bibinfo {title} {Terabit free-space data transmission employing orbital angular momentum multiplexing},\ }\href {https://doi.org/10.1038/nphoton.2012.138} {\bibfield  {journal} {\bibinfo  {journal} {Nat. Photon.}\ }\textbf {\bibinfo {volume} {6}},\ \bibinfo {pages} {488} (\bibinfo {year}
  {2012})}\BibitemShut {NoStop}%
\bibitem [{\citenamefont {Vallone}\ \emph {et~al.}(2014)\citenamefont {Vallone}, \citenamefont {D'Ambrosio}, \citenamefont {Sponselli}, \citenamefont {Slussarenko}, \citenamefont {Marrucci}, \citenamefont {Sciarrino},\ and\ \citenamefont {Villoresi}}]{Vallone2014}%
  \BibitemOpen
  \bibfield  {author} {\bibinfo {author} {\bibfnamefont {G.}~\bibnamefont {Vallone}}, \bibinfo {author} {\bibfnamefont {V.}~\bibnamefont {D'Ambrosio}}, \bibinfo {author} {\bibfnamefont {A.}~\bibnamefont {Sponselli}}, \bibinfo {author} {\bibfnamefont {S.}~\bibnamefont {Slussarenko}}, \bibinfo {author} {\bibfnamefont {L.}~\bibnamefont {Marrucci}}, \bibinfo {author} {\bibfnamefont {F.}~\bibnamefont {Sciarrino}},\ and\ \bibinfo {author} {\bibfnamefont {P.}~\bibnamefont {Villoresi}},\ }\bibfield  {title} {\bibinfo {title} {Free-space quantum key distribution by rotation-invariant twisted photons},\ }\href {https://doi.org/10.1103/PhysRevLett.113.060503} {\bibfield  {journal} {\bibinfo  {journal} {Phys. Rev. Lett.}\ }\textbf {\bibinfo {volume} {113}},\ \bibinfo {pages} {060503} (\bibinfo {year} {2014})}\BibitemShut {NoStop}%
\bibitem [{\citenamefont {Krenn}\ \emph {et~al.}(2014)\citenamefont {Krenn}, \citenamefont {Fickler}, \citenamefont {Fink}, \citenamefont {Handsteiner}, \citenamefont {Malik}, \citenamefont {Scheidl}, \citenamefont {Ursin},\ and\ \citenamefont {Zeilinger}}]{Krenn2014}%
  \BibitemOpen
  \bibfield  {author} {\bibinfo {author} {\bibfnamefont {M.}~\bibnamefont {Krenn}}, \bibinfo {author} {\bibfnamefont {R.}~\bibnamefont {Fickler}}, \bibinfo {author} {\bibfnamefont {M.}~\bibnamefont {Fink}}, \bibinfo {author} {\bibfnamefont {J.}~\bibnamefont {Handsteiner}}, \bibinfo {author} {\bibfnamefont {M.}~\bibnamefont {Malik}}, \bibinfo {author} {\bibfnamefont {T.}~\bibnamefont {Scheidl}}, \bibinfo {author} {\bibfnamefont {R.}~\bibnamefont {Ursin}},\ and\ \bibinfo {author} {\bibfnamefont {A.}~\bibnamefont {Zeilinger}},\ }\bibfield  {title} {\bibinfo {title} {Communication with spatially modulated light through turbulent air across vienna},\ }\href {https://doi.org/10.1088/1367-2630/16/11/113028} {\bibfield  {journal} {\bibinfo  {journal} {New Journal of Physics}\ }\textbf {\bibinfo {volume} {16}},\ \bibinfo {pages} {113028} (\bibinfo {year} {2014})}\BibitemShut {NoStop}%
\bibitem [{\citenamefont {Mirhosseini}\ \emph {et~al.}(2015)\citenamefont {Mirhosseini}, \citenamefont {Magaña-Loaiza}, \citenamefont {O’Sullivan}, \citenamefont {Rodenburg}, \citenamefont {Malik}, \citenamefont {Lavery}, \citenamefont {Padgett}, \citenamefont {Gauthier},\ and\ \citenamefont {Boyd}}]{Mirhosseini:15}%
  \BibitemOpen
  \bibfield  {author} {\bibinfo {author} {\bibfnamefont {M.}~\bibnamefont {Mirhosseini}}, \bibinfo {author} {\bibfnamefont {O.~S.}\ \bibnamefont {Magaña-Loaiza}}, \bibinfo {author} {\bibfnamefont {M.~N.}\ \bibnamefont {O’Sullivan}}, \bibinfo {author} {\bibfnamefont {B.}~\bibnamefont {Rodenburg}}, \bibinfo {author} {\bibfnamefont {M.}~\bibnamefont {Malik}}, \bibinfo {author} {\bibfnamefont {M.~P.~J.}\ \bibnamefont {Lavery}}, \bibinfo {author} {\bibfnamefont {M.~J.}\ \bibnamefont {Padgett}}, \bibinfo {author} {\bibfnamefont {D.~J.}\ \bibnamefont {Gauthier}},\ and\ \bibinfo {author} {\bibfnamefont {R.~W.}\ \bibnamefont {Boyd}},\ }\bibfield  {title} {\bibinfo {title} {High-dimensional quantum cryptography with twisted light},\ }\href {https://doi.org/10.1088/1367-2630/17/3/033033} {\bibfield  {journal} {\bibinfo  {journal} {New Journal of Phys.}\ }\textbf {\bibinfo {volume} {17}},\ \bibinfo {pages} {033033} (\bibinfo {year} {2015})}\BibitemShut {NoStop}%
\bibitem [{\citenamefont {Sit}\ \emph {et~al.}(2017)\citenamefont {Sit}, \citenamefont {Bouchard}, \citenamefont {Fickler}, \citenamefont {Gagnon-Bischoff}, \citenamefont {Larocque}, \citenamefont {Heshami}, \citenamefont {Elser}, \citenamefont {Peuntinger}, \citenamefont {G\"{u}nthner}, \citenamefont {Heim}, \citenamefont {Marquardt}, \citenamefont {Leuchs}, \citenamefont {Boyd},\ and\ \citenamefont {Karimi}}]{Sit2017}%
  \BibitemOpen
  \bibfield  {author} {\bibinfo {author} {\bibfnamefont {A.}~\bibnamefont {Sit}}, \bibinfo {author} {\bibfnamefont {F.}~\bibnamefont {Bouchard}}, \bibinfo {author} {\bibfnamefont {R.}~\bibnamefont {Fickler}}, \bibinfo {author} {\bibfnamefont {J.}~\bibnamefont {Gagnon-Bischoff}}, \bibinfo {author} {\bibfnamefont {H.}~\bibnamefont {Larocque}}, \bibinfo {author} {\bibfnamefont {K.}~\bibnamefont {Heshami}}, \bibinfo {author} {\bibfnamefont {D.}~\bibnamefont {Elser}}, \bibinfo {author} {\bibfnamefont {C.}~\bibnamefont {Peuntinger}}, \bibinfo {author} {\bibfnamefont {K.}~\bibnamefont {G\"{u}nthner}}, \bibinfo {author} {\bibfnamefont {B.}~\bibnamefont {Heim}}, \bibinfo {author} {\bibfnamefont {C.}~\bibnamefont {Marquardt}}, \bibinfo {author} {\bibfnamefont {G.}~\bibnamefont {Leuchs}}, \bibinfo {author} {\bibfnamefont {R.~W.}\ \bibnamefont {Boyd}},\ and\ \bibinfo {author} {\bibfnamefont {E.}~\bibnamefont {Karimi}},\ }\bibfield  {title} {\bibinfo {title} {High-dimensional intracity quantum cryptography with structured
  photons},\ }\href {https://doi.org/10.1364/OPTICA.4.001006} {\bibfield  {journal} {\bibinfo  {journal} {Optica}\ }\textbf {\bibinfo {volume} {4}},\ \bibinfo {pages} {1006} (\bibinfo {year} {2017})}\BibitemShut {NoStop}%
\bibitem [{\citenamefont {Liu}\ \emph {et~al.}(2020)\citenamefont {Liu}, \citenamefont {Lou},\ and\ \citenamefont {Jing}}]{Liu2020}%
  \BibitemOpen
  \bibfield  {author} {\bibinfo {author} {\bibfnamefont {S.}~\bibnamefont {Liu}}, \bibinfo {author} {\bibfnamefont {Y.}~\bibnamefont {Lou}},\ and\ \bibinfo {author} {\bibfnamefont {J.}~\bibnamefont {Jing}},\ }\bibfield  {title} {\bibinfo {title} {Orbital angular momentum multiplexed deterministic all-optical quantum teleportation},\ }\href {https://doi.org/10.1038/s41467-020-17616-4} {\bibfield  {journal} {\bibinfo  {journal} {Nature Communications}\ }\textbf {\bibinfo {volume} {11}},\ \bibinfo {pages} {3875} (\bibinfo {year} {2020})}\BibitemShut {NoStop}%
\bibitem [{\citenamefont {Choi}\ \emph {et~al.}(2012)\citenamefont {Choi}, \citenamefont {Yoon}, \citenamefont {Kim}, \citenamefont {Yang}, \citenamefont {Fang-Yen}, \citenamefont {Dasari}, \citenamefont {Lee},\ and\ \citenamefont {Choi}}]{Choi2012}%
  \BibitemOpen
  \bibfield  {author} {\bibinfo {author} {\bibfnamefont {Y.}~\bibnamefont {Choi}}, \bibinfo {author} {\bibfnamefont {C.}~\bibnamefont {Yoon}}, \bibinfo {author} {\bibfnamefont {M.}~\bibnamefont {Kim}}, \bibinfo {author} {\bibfnamefont {T.~D.}\ \bibnamefont {Yang}}, \bibinfo {author} {\bibfnamefont {C.}~\bibnamefont {Fang-Yen}}, \bibinfo {author} {\bibfnamefont {R.~R.}\ \bibnamefont {Dasari}}, \bibinfo {author} {\bibfnamefont {K.~J.}\ \bibnamefont {Lee}},\ and\ \bibinfo {author} {\bibfnamefont {W.}~\bibnamefont {Choi}},\ }\bibfield  {title} {\bibinfo {title} {Scanner-free and wide-field endoscopic imaging by using a single multimode optical fiber},\ }\href {https://doi.org/10.1103/PhysRevLett.109.203901} {\bibfield  {journal} {\bibinfo  {journal} {Phys. Rev. Lett.}\ }\textbf {\bibinfo {volume} {109}},\ \bibinfo {pages} {203901} (\bibinfo {year} {2012})}\BibitemShut {NoStop}%
\bibitem [{\citenamefont {Caramazza}\ \emph {et~al.}(2019)\citenamefont {Caramazza}, \citenamefont {Moran}, \citenamefont {Murray-Smith},\ and\ \citenamefont {Faccio}}]{Caramazza2019}%
  \BibitemOpen
  \bibfield  {author} {\bibinfo {author} {\bibfnamefont {P.}~\bibnamefont {Caramazza}}, \bibinfo {author} {\bibfnamefont {O.}~\bibnamefont {Moran}}, \bibinfo {author} {\bibfnamefont {R.}~\bibnamefont {Murray-Smith}},\ and\ \bibinfo {author} {\bibfnamefont {D.}~\bibnamefont {Faccio}},\ }\bibfield  {title} {\bibinfo {title} {Transmission of natural scene images through a multimode fibre},\ }\href {https://doi.org/10.1038/s41467-019-10057-8} {\bibfield  {journal} {\bibinfo  {journal} {Nature Communications}\ }\textbf {\bibinfo {volume} {10}},\ \bibinfo {pages} {2029} (\bibinfo {year} {2019})}\BibitemShut {NoStop}%
\bibitem [{\citenamefont {Richardson}\ \emph {et~al.}(2013)\citenamefont {Richardson}, \citenamefont {Fini},\ and\ \citenamefont {Nelson}}]{Richardson2013}%
  \BibitemOpen
  \bibfield  {author} {\bibinfo {author} {\bibfnamefont {D.~J.}\ \bibnamefont {Richardson}}, \bibinfo {author} {\bibfnamefont {J.~M.}\ \bibnamefont {Fini}},\ and\ \bibinfo {author} {\bibfnamefont {L.~E.}\ \bibnamefont {Nelson}},\ }\bibfield  {title} {\bibinfo {title} {Space-division multiplexing in optical fibres},\ }\href {https://doi.org/10.1038/nphoton.2013.94} {\bibfield  {journal} {\bibinfo  {journal} {Nat. Photon.}\ }\textbf {\bibinfo {volume} {7}},\ \bibinfo {pages} {354} (\bibinfo {year} {2013})}\BibitemShut {NoStop}%
\bibitem [{\citenamefont {Xavier}\ and\ \citenamefont {Lima}(2020)}]{Xavier:20}%
  \BibitemOpen
  \bibfield  {author} {\bibinfo {author} {\bibfnamefont {G.~B.}\ \bibnamefont {Xavier}}\ and\ \bibinfo {author} {\bibfnamefont {G.}~\bibnamefont {Lima}},\ }\bibfield  {title} {\bibinfo {title} {Quantum information processing with space-division multiplexing optical fibres},\ }\href {https://doi.org/https://doi.org/10.1038/s42005-019-0269-7} {\bibfield  {journal} {\bibinfo  {journal} {Commun. Phys.}\ }\textbf {\bibinfo {volume} {3}},\ \bibinfo {pages} {9} (\bibinfo {year} {2020})}\BibitemShut {NoStop}%
\bibitem [{\citenamefont {Cao}\ \emph {et~al.}(2020)\citenamefont {Cao}, \citenamefont {Gao}, \citenamefont {Zhang}, \citenamefont {Wang}, \citenamefont {He}, \citenamefont {Liu}, \citenamefont {Zhou}, \citenamefont {Chen}, \citenamefont {Li}, \citenamefont {Yu}, \citenamefont {Romero}, \citenamefont {Huang}, \citenamefont {Li},\ and\ \citenamefont {Guo}}]{Cao2020}%
  \BibitemOpen
  \bibfield  {author} {\bibinfo {author} {\bibfnamefont {H.}~\bibnamefont {Cao}}, \bibinfo {author} {\bibfnamefont {S.-C.}\ \bibnamefont {Gao}}, \bibinfo {author} {\bibfnamefont {C.}~\bibnamefont {Zhang}}, \bibinfo {author} {\bibfnamefont {J.}~\bibnamefont {Wang}}, \bibinfo {author} {\bibfnamefont {D.-Y.}\ \bibnamefont {He}}, \bibinfo {author} {\bibfnamefont {B.-H.}\ \bibnamefont {Liu}}, \bibinfo {author} {\bibfnamefont {Z.-W.}\ \bibnamefont {Zhou}}, \bibinfo {author} {\bibfnamefont {Y.-J.}\ \bibnamefont {Chen}}, \bibinfo {author} {\bibfnamefont {Z.-H.}\ \bibnamefont {Li}}, \bibinfo {author} {\bibfnamefont {S.-Y.}\ \bibnamefont {Yu}}, \bibinfo {author} {\bibfnamefont {J.}~\bibnamefont {Romero}}, \bibinfo {author} {\bibfnamefont {Y.-F.}\ \bibnamefont {Huang}}, \bibinfo {author} {\bibfnamefont {C.-F.}\ \bibnamefont {Li}},\ and\ \bibinfo {author} {\bibfnamefont {G.-C.}\ \bibnamefont {Guo}},\ }\bibfield  {title} {\bibinfo {title} {Distribution of high-dimensional orbital angular momentum entanglement over a 1 km
  few-mode fiber},\ }\href {https://doi.org/10.1364/OPTICA.381403} {\bibfield  {journal} {\bibinfo  {journal} {Optica}\ }\textbf {\bibinfo {volume} {7}},\ \bibinfo {pages} {232} (\bibinfo {year} {2020})}\BibitemShut {NoStop}%
\bibitem [{\citenamefont {Alarc\'on}\ \emph {et~al.}(2021)\citenamefont {Alarc\'on}, \citenamefont {Argillander}, \citenamefont {Lima},\ and\ \citenamefont {Xavier}}]{Alarcon2021}%
  \BibitemOpen
  \bibfield  {author} {\bibinfo {author} {\bibfnamefont {A.}~\bibnamefont {Alarc\'on}}, \bibinfo {author} {\bibfnamefont {J.}~\bibnamefont {Argillander}}, \bibinfo {author} {\bibfnamefont {G.}~\bibnamefont {Lima}},\ and\ \bibinfo {author} {\bibfnamefont {G.}~\bibnamefont {Xavier}},\ }\bibfield  {title} {\bibinfo {title} {Few-mode-fiber technology fine-tunes losses in quantum communication systems},\ }\href {https://doi.org/10.1103/PhysRevApplied.16.034018} {\bibfield  {journal} {\bibinfo  {journal} {Phys. Rev. Appl.}\ }\textbf {\bibinfo {volume} {16}},\ \bibinfo {pages} {034018} (\bibinfo {year} {2021})}\BibitemShut {NoStop}%
\bibitem [{\citenamefont {Birks}\ \emph {et~al.}(2015)\citenamefont {Birks}, \citenamefont {Gris-S\'{a}nchez}, \citenamefont {Yerolatsitis}, \citenamefont {Leon-Saval},\ and\ \citenamefont {Thomson}}]{Birks:15}%
  \BibitemOpen
  \bibfield  {author} {\bibinfo {author} {\bibfnamefont {T.~A.}\ \bibnamefont {Birks}}, \bibinfo {author} {\bibfnamefont {I.}~\bibnamefont {Gris-S\'{a}nchez}}, \bibinfo {author} {\bibfnamefont {S.}~\bibnamefont {Yerolatsitis}}, \bibinfo {author} {\bibfnamefont {S.~G.}\ \bibnamefont {Leon-Saval}},\ and\ \bibinfo {author} {\bibfnamefont {R.~R.}\ \bibnamefont {Thomson}},\ }\bibfield  {title} {\bibinfo {title} {The photonic lantern},\ }\href {https://doi.org/10.1364/AOP.7.000107} {\bibfield  {journal} {\bibinfo  {journal} {Adv. Opt. Photon.}\ }\textbf {\bibinfo {volume} {7}},\ \bibinfo {pages} {107} (\bibinfo {year} {2015})}\BibitemShut {NoStop}%
\bibitem [{\citenamefont {Alarc\'{o}n}\ \emph {et~al.}(2023)\citenamefont {Alarc\'{o}n}, \citenamefont {Argillander}, \citenamefont {Spegel-Lexne},\ and\ \citenamefont {Xavier}}]{Alarcon2023_2}%
  \BibitemOpen
  \bibfield  {author} {\bibinfo {author} {\bibfnamefont {A.}~\bibnamefont {Alarc\'{o}n}}, \bibinfo {author} {\bibfnamefont {J.}~\bibnamefont {Argillander}}, \bibinfo {author} {\bibfnamefont {D.}~\bibnamefont {Spegel-Lexne}},\ and\ \bibinfo {author} {\bibfnamefont {G.~B.}\ \bibnamefont {Xavier}},\ }\bibfield  {title} {\bibinfo {title} {Dynamic generation of photonic spatial quantum states with an all-fiber platform},\ }\href {https://doi.org/10.1364/OE.481974} {\bibfield  {journal} {\bibinfo  {journal} {Opt. Express}\ }\textbf {\bibinfo {volume} {31}},\ \bibinfo {pages} {10673} (\bibinfo {year} {2023})}\BibitemShut {NoStop}%
\bibitem [{\citenamefont {Alarc{\'o}n}\ \emph {et~al.}(2023)\citenamefont {Alarc{\'o}n}, \citenamefont {G{\'o}mez}, \citenamefont {Spegel-Lexne}, \citenamefont {Argillander}, \citenamefont {Cari{\~n}e}, \citenamefont {Ca{\~n}as}, \citenamefont {Lima},\ and\ \citenamefont {Xavier}}]{Alarcon2023}%
  \BibitemOpen
  \bibfield  {author} {\bibinfo {author} {\bibfnamefont {A.}~\bibnamefont {Alarc{\'o}n}}, \bibinfo {author} {\bibfnamefont {S.}~\bibnamefont {G{\'o}mez}}, \bibinfo {author} {\bibfnamefont {D.}~\bibnamefont {Spegel-Lexne}}, \bibinfo {author} {\bibfnamefont {J.}~\bibnamefont {Argillander}}, \bibinfo {author} {\bibfnamefont {J.}~\bibnamefont {Cari{\~n}e}}, \bibinfo {author} {\bibfnamefont {G.}~\bibnamefont {Ca{\~n}as}}, \bibinfo {author} {\bibfnamefont {G.}~\bibnamefont {Lima}},\ and\ \bibinfo {author} {\bibfnamefont {G.~B.}\ \bibnamefont {Xavier}},\ }\bibfield  {title} {\bibinfo {title} {All-in-fiber dynamically reconfigurable orbital angular momentum mode sorting},\ }\href {https://doi.org/10.1021/acsphotonics.3c00825} {\bibfield  {journal} {\bibinfo  {journal} {ACS Photonics}\ }\textbf {\bibinfo {volume} {10}},\ \bibinfo {pages} {3700} (\bibinfo {year} {2023})}\BibitemShut {NoStop}%
\bibitem [{\citenamefont {Argillander}\ \emph {et~al.}(2022)\citenamefont {Argillander}, \citenamefont {Alarc{\'o}n},\ and\ \citenamefont {Xavier}}]{Argillander_2022}%
  \BibitemOpen
  \bibfield  {author} {\bibinfo {author} {\bibfnamefont {J.}~\bibnamefont {Argillander}}, \bibinfo {author} {\bibfnamefont {A.}~\bibnamefont {Alarc{\'o}n}},\ and\ \bibinfo {author} {\bibfnamefont {G.~B.}\ \bibnamefont {Xavier}},\ }\bibfield  {title} {\bibinfo {title} {A tunable quantum random number generator based on a fiber-optical sagnac interferometer},\ }\href {https://doi.org/10.1088/2040-8986/ac68f4} {\bibfield  {journal} {\bibinfo  {journal} {Journal of Optics}\ }\textbf {\bibinfo {volume} {24}},\ \bibinfo {pages} {064010} (\bibinfo {year} {2022})}\BibitemShut {NoStop}%
\bibitem [{\citenamefont {Qi}\ \emph {et~al.}(2006)\citenamefont {Qi}, \citenamefont {Fung}, \citenamefont {Lo},\ and\ \citenamefont {Ma}}]{Lo_timeshift}%
  \BibitemOpen
  \bibfield  {author} {\bibinfo {author} {\bibfnamefont {B.}~\bibnamefont {Qi}}, \bibinfo {author} {\bibfnamefont {C.-H.~F.}\ \bibnamefont {Fung}}, \bibinfo {author} {\bibfnamefont {H.-K.}\ \bibnamefont {Lo}},\ and\ \bibinfo {author} {\bibfnamefont {X.}~\bibnamefont {Ma}},\ }\bibfield  {title} {\bibinfo {title} {Time-shift attack in practical quantum cryptosystems},\ }\href@noop {} {\bibfield  {journal} {\bibinfo  {journal} {Quantum Information and Computation}\ }\textbf {\bibinfo {volume} {7}},\ \bibinfo {pages} {073} (\bibinfo {year} {2006})}\BibitemShut {NoStop}%
\end{thebibliography}%


\end{document}